\documentclass{aa}  

\usepackage{graphicx}
\usepackage{txfonts}
\usepackage[colorlinks,linkcolor=blue,anchorcolor=blue,citecolor=blue]{hyperref}
\usepackage{graphicx}
\usepackage{subcaption}
\usepackage{longtable}
\usepackage{placeins}

\usepackage{tikz}

\usepackage{xcolor, soul}
\sethlcolor{green}
\begin{document}

   \title{The population of tidal disruption events discovered with eROSITA}

   \author{I. Grotova\inst{\ref{inst1}}
          \and A. Rau\inst{\ref{inst1}}
          \and P. Baldini\inst{\ref{inst1}}
          \and A. J. Goodwin\inst{\ref{inst2}}
          \and Z. Liu\inst{\ref{inst1}}
           \and A. Merloni\inst{\ref{inst1}}
          \and M. Salvato\inst{\ref{inst1}}
          \and G. E. Anderson\inst{\ref{inst2}}
          \and R. Arcodia\inst{\ref{inst9}}
          \and J. Buchner\inst{\ref{inst1}}
          \and M. Krumpe\inst{\ref{inst6}}
          \and A. Malyali\inst{\ref{inst1}}
          \and M. Masterson\inst{\ref{inst9}}
          \and J. C. A. Miller-Jones\inst{\ref{inst2}}  
          \and K. Nandra\inst{\ref{inst1}}
          \and R. Shirley\inst{\ref{inst1}}    
        }
   \institute{Max-Planck-Institut f\"ur extraterrestrische Physik, Giessenbachstrasse 1, 85748 Garching, Germany\label{inst1}\\\email{grotova@mpe.mpg.de}
   \and Leibniz-Institut f\"ur Astrophysik Potsdam, An der Sternwarte 16, 14482 Potsdam, Germany\label{inst6}
   \and International Center for Radio Astronomy Research, Curtin University, GPO Box U1987, Perth, WA 6845, Australia\label{inst2}
   \and MIT Kavli Institute for Astrophysics and Space Research, 70 Vassar Street, Cambridge, MA 02139, USA \label{inst9}
    }

   \date{Received January 3, 2025; accepted March 13, 2025}
 
  \abstract
   {}
   {The { \it Spectrum Roentgen Gamma} (SRG) eROSITA all-sky survey marks the beginning of the data-rich era by conducting population studies of tidal disruption events (TDEs) and other rare X-ray transients. This paper presents a systematic study of X-ray-selected canonical TDEs discovered in the western Galactic hemisphere of the first two eROSITA all-sky surveys (eRASS1 and eRASS2) performed between Dec 2019 and Dec 2020.}
   {We compiled a TDE sample from the catalog of eROSITA’s extragalactic transients and variables eRO-ExTra, which includes X-ray sources with a variability significance and fractional amplitude over four between eRASS1 and eRASS2, not associated with known active galactic nuclei (AGNs). Each X-ray source is associated with an optical counterpart from the Legacy Survey DR10 (LS10). Canonical TDEs were selected based on their X-ray light-curve properties (single flare or decline), soft X-ray spectra ($\Gamma>3$), and the absence of archival X-ray variability and AGN signatures in their host photometry and spectroscopy.}
   {We present 31 X-ray-selected TDE candidates associated with optical counterparts with redshifts of $0.02<z<0.34$ and luminosities of $\mathrm{5.7\times10^{41}<L_{X}<5.3\times10^{44} ergs^{-1}}$ in the $0.2 - 6.0$ keV rest frame. The sample contains 30 canonical TDEs and one off-nuclear TDE candidate. The X-ray luminosity function derived from this sample is best fit by a double power law with a luminosity break at $\mathrm{10^{44}ergs^{-1}}$, corresponding to the Eddington-limiting prediction. The result is in agreement with previous observational and theoretical estimates. This corresponds to a TDE volumetric rate of $\mathrm{ (2.3^{+1.2}_{-0.9})\times10^{-7}\,Mpc^{-3} yr^{-1}}$ ($\approx1.2\times 10^{-5}$ events per galaxy per year). The TDE host galaxies show a green-valley overdensity, as was previously found in X-ray and optical studies. In addition, 20\,\%, 30\,\%, and 15\,\% of our X-ray-selected sample exhibit flares in the optical, mid-infrared (mid-IR), or radio bands, respectively. We discuss the differences between X-ray, optical, and mid-IR TDE populations and the origins of multiwavelength flares in the context of the obscuring envelope and stream-stream collision models. Finally, we highlight TDE subpopulations that are not included in the canonical sample and should be explored in the future.

}
   {}

   \keywords{transients -- X-ray
               }

   \maketitle
%

\section{Introduction}
A star passing within a tidal radius of a massive black hole will be disrupted by gravitational forces. In such a tidal disruption event (TDE; \citealt{1975Natur.254..295H,1988Natur.333..523R}), the bound portion of stellar debris will eventually be accreted onto the black hole (BH), resulting in a bright electromagnetic flare lasting from weeks to years. TDEs provide unique insight into the otherwise inaccessible environments of quiescent BHs, their demographics, and dynamical accretion processes.

 The first TDE candidates were discovered in X-rays by the ROSAT all-sky survey (e.g., \citealt{1996A&A...309L..35B,1999A&A...350L..31G,komossa1999discovery,2000A&A...362L..25G}). They showed large-amplitude soft X-ray flares ($\approx \rm{few} \times 10^5 K$) in the centers of quiescent galaxies and a decline ($\mathrm{\propto t^{-5/3}}$; \citealt{komossa2015tidal}) on timescales ranging from months to years. Further X-ray-selected TDE candidates came from the {\it XMM-Newton} Slew survey, {\it Chandra}, and the \textit{Neil Gehrels} {\it Swift} Observatory. A comprehensive review of 21 X-ray-discovered TDE candidates reported in the literature by mid-2019 is provided in \citet{2020SSRv..216...85S}. Launched in 2019, the extended Roentgen Survey with an Imaging Telescope Array (eROSITA;  \citealt{predehl_2021}) on board the Russian-German Spektr-RG mission \citep{2021A&A...656A.132S} has revealed numerous canonical TDEs and uncovered more exotic TDE scenarios \citep{Malyali_AT2019avd,Malyali_2023,malyali22dsb}. For example, in the German half of the eROSITA sky (eROSITA\_DE; $359.9442 ^{\circ}>l> 179.9442 ^{\circ}$), repeating and partial TDEs (pTDEs) exhibiting repeating flares on timescales ranging from weeks (e.g., eRASSt J045650.3-203750; \citealt{Liu_2023}, \citealt{Liu_2024}) to decades (e.g., RX J133157.6-324319.7; \citealt{malyali_rep}) were discovered. The eROSITA all-sky coverage and higher sensitivity also enabled us to perform the first systematic X-ray TDE population studies. A study of a sample of 13 TDE candidates found between Jun 2020 and Dec 2020 in the eastern Galactic hemisphere (eROSITA\_RU; $-0.0558^{\circ}< l <179.9442^{\circ}$) data of the eROSITA all-sky survey was presented in \cite{Sazonov_2021}.

Thanks to the advent of large wide-field optical surveys (e.g., \textit{Zwicky} Transient Facility; \citealt{2019PASP..131a8002B},  ASAS-SN; \citealt{2014ApJ...788...48S,2017PASP..129j4502K}), the discovery rate of TDEs in the optical band has significantly increased in recent years (see a review of 56 TDE candidates in \citealt{Gezari_2021}), enabling the first population studies \citep{vanVelzen_2021, yao2023,guolo2024systematic}. Population studies were also performed in mid-infrared (mid-IR; \citealt{masterson2024}) and radio bands \citep{2024ApJ...973..104D,goodwin2025}. The growing sample of events demonstrated that the underlying population of TDEs is very heterogeneous; this is supported by the differences in the predicted TDE luminosity functions and rates. For example, not all X-ray-selected TDEs show optical or infrared (IR) emission and vice versa; many show notable deviations from the expected, monotonically decaying light-curve evolution, including combinations of missing or altered multiwavelength signatures \citep{van_Velzen_2020,masterson2024}. The apparent dichotomy in the multiwavelength emission of TDEs, reflected in differing characteristic radii and temperatures, can be attributed to distinct physical processes and emission regions. The X-ray emission, consistent with thermal emission, is expected to trace the formation of the accretion disk around the supermassive black hole (SMBH). Conversely, the optical emission is thought to arise from collisions of the debris stream at large radii during the circularization process \citep{piran2015,Jiang2016, bonnerot2017}. Alternatively, optical and IR emission may come from the reprocessing of the X-ray emission by an optically thick envelope \citep{Roth_2016,Dai2018,2022ApJ...937L..28T}. In the latter scenario, the observed differences between the optical and X-ray emission properties are explained by the observer’s viewing angle \citep{Dai2018,parkinson22} and patchy obscuration \citep{van_Velzen_2020}. It is an open question as to which scenario is preferential for describing observational properties of TDEs and whether several processes co-exist during different stages of TDE evolution. Moreover, recent samples of X-ray- \citep{Sazonov_2021} and optically \citep{yao2023} selected TDEs uncovered lower TDE rates ($\mathrm{\approx10^{-5} year^{-1}galaxy^{-1}}$) compared to the theoretical rates predicted by loss-cone dynamics \citep{2011MNRAS.418.1308B, 2015ApJ...814..141M,2016MNRAS.455..859S} on the order of $\mathrm{10^{-4}year^{-1}galaxy^{-1}}$. Population studies involving systematic all-sky studies will shed more light on the missing TDE problem, whether it arises from observational caveats, selection biases of TDEs in different energy bands, or the physical interpretation of multiwavelength TDE emission. 

Another puzzle about the TDE host galaxies is that optical and X-ray TDEs are mostly identified in the "green valley" of the galaxy color-mass space (e.g., \citealt{2017ApJ...850...22L,Hammerstein_2021,Sazonov_2021, yao2023}), which is associated with galaxies between star-forming ("blue") and passive ("red") states \citep{2007ApJS..173..342M}. Whereas the larger TDE rate in the green-valley galaxies may be explained by an efficient loss-cone filling after a recent star formation period or a galaxy merger \citep{2020SSRv..216...32F,Hammerstein_2021}, it is still unclear if the overpopulation is biased against redder hosts wherein the TDEs can be obscured by dust \citep{Roth_2021}. 

In this work, we present a sample of the 31 most prominent TDE candidates selected from the first two eROSITA all-sky surveys (eRASS) in the western Galactic hemisphere. The presented sample is the largest systematically X-ray-selected TDE sample to date due to the unprecedented sensitivity, six-month cadence, and all-sky coverage of eROSITA. The sample includes 30 canonical TDEs and one off-nuclear TDE candidate. We provide a statistical study of the population using in-depth X-ray analysis as well as multiwavelength data in mid-IR, optical, and radio bands. The study focuses on improving existing X-ray TDE rate estimates, inspecting the TDE host population and BH mass distribution and exploring a multiband view of TDEs on a timescale of years.  

The paper is structured as follows. In Sect.~\ref{sec:selection}, we outline the selection of the TDE sample; in Sect.~\ref{sec:final_sample} we give an overview of the final sample, its X-ray properties, and optical counterparts; in Sect.~\ref{sec:multivar} we explore the multiwavelength variability of the selected TDEs; and in Sect.~\ref{sec:sed} we analyze the host galaxy properties of the sample. Finally, we provide a discussion in Sect.~\ref{sec:discussion}. The computation of the TDE X-ray luminosity function and rates are given in Sect.~\ref{sec:xlf}; and we conclude with a summary in Sect.~\ref{sec:summary}.
   
We adopted a flat $\mathrm{\Lambda}$CDM cosmology throughout this work, with $H_{0}=67.7$\,km s$^{-1}$ Mpc$^{-1}$ and $\Omega_m=0.309$ \citep{2016A&A...594A..13P}. The uncertainties quoted in
the paper are based on 68\% confidence intervals, and upper limits (ULs) are reported at $3\sigma$, unless stated otherwise. 

\section{Selection of TDE candidates}
\label{sec:selection}

X-ray emission is a direct probe of BH accretion and, therefore, is the most robust way of identifying accretion-driven flares. This study focuses on canonical X-ray TDEs, that is events characterized by an X-ray light curve showing a single flare lasting months to years with a soft ($\Gamma\gtrsim  3$) spectrum. The parent sample is the catalog of extragalactic, non-AGN X-ray transients and variables eRO-ExTra\footnote{\url{https://cdsarc.cds.unistra.fr/viz-bin/cat/J/A+A/693/A62}} \citep{Grotova_2025}. The following subsections detail the TDE selection process based on X-ray light-curve and spectral properties.

\subsection{Parent catalog: eRO-ExTra}

 eRO-ExTra is a systematically selected catalog of 304 X-ray transients and variables discovered in the first two eROSITA\_DE all-sky surveys \citep[][eRASS1 and eRASS2; Nov 2019--Nov 2020]{Merloni_2024} and contained within the Legacy survey DR10 (LS10\footnote{\url{https://www.legacysurvey.org/dr10/}}; \citealt{2019AJ....157..168D}) footprint. Starting from more than two thousand sources, which vary with significance and fractional amplitude larger than four between eRASS1 and eRASS2, eRO-ExTra was thoroughly cleaned from contaminants, such as stars and traditional active galactic nuclei (AGNs). The eRO-ExTra catalog made use of available archival photometric and spectroscopic data to exclude AGN contaminants. In particular, a mid-IR cut of $\mathrm{W1-W2<0.8\,mag_{ Vega}}$ \citep{stern2012} was applied to exclude typical AGNs based on their mid-IR color indicative of hot dust emission. Moreover, available pre-flare archival spectra were inspected, and sources with typical AGN emission lines were removed. Other applied criteria included a parallax cut; visual inspection and utilization of previous classification from the literature; and the analysis of X-ray data from other missions. Namely, archival X-ray data helped distinguish new transients from long-term X-ray variables such as AGNs. The eRO-ExTra catalog has been cleaned from sources showing significant archival X-ray detections at the level of eROSITA peaks using {\it XMM-Newton} Slew Survey \citep{Saxton_2008_xmmslew}, {\it Swift} \citep{swiftxrt}, and {\it ROSAT} \citep{voges1999rosat} archival data.

All eRO-ExTra sources were associated with optical LS10 counterparts using a Bayesian algorithm for cross-matching multiple catalogs called NWAY \citep{Salvato2018,Salvato_2022}. Based on the assigned NWAY counterpart probability (\texttt{p\_any}>0.17), more than 90\% of the counterparts are considered reliable. Moreover, reliable spectroscopic or photometric redshifts \citep{2024A&A...690A.365S} are provided for more than 80\% of the sample.
The systematic selection of eRO-ExTra and provided long-term eROSITA light-curve classification and X-ray spectral modeling at the peak epoch allowed us to identify strong canonical TDE candidates. 

\begin{figure*}
\centering
    \includegraphics[width=0.9\linewidth]{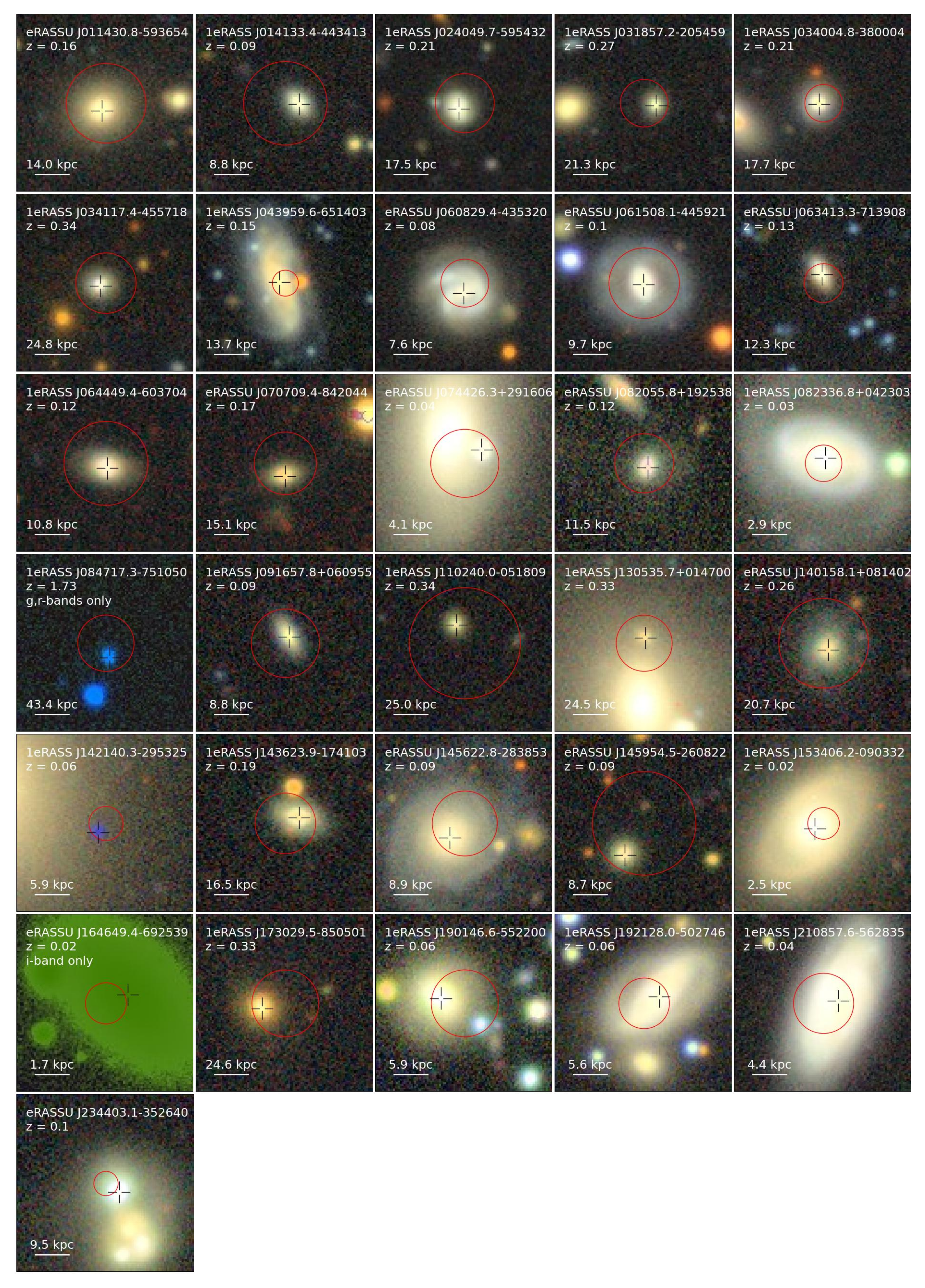}  
\caption{LS10 {\it griz}-color images ($\mathrm{100\times 100}$ pixels) of the host galaxies of the X-ray TDE sample. Black crosses denote the LS10 counterpart positions, and red circles show the eROSITA X-ray positions with their $3\sigma$ errors. Images for J1646 and J0847 were only taken in {\it i} and {\it g,r} bands, respectively. The counterpart of J1421 appears blue, since it was only detected in the {\it g} band (see Appendix~\ref{sec:j1421}). }
\label{fig:ls10_cutouts}
\end{figure*}

\subsection{eROSITA light curves}

X-ray light curves allowed us to further differentiate among TDEs, variable AGNs, and AGN ingition events. Canonically, the X-ray light curve of a TDE shows a sharp rise, peaking within a few weeks to a few months after the disruption event, followed by a power-law decay over several months to years with an index of $-5/3$ \citep{1989ApJ...346L..13E}; the typical X-ray peak luminosities are between $10^{42}$ and a few times $10^{44}$ $\mathrm{erg s^{-1}}$ \citep{2020SSRv..216...85S}. However, the fall-back rate can be influenced by various factors such as the internal structure and spin of a star, the BH spin, and the impact parameter of the star's orbit \citep{Gezari_2021}. The assumption of a full disruption with debris following Keplerian orbits will also be violated in pTDEs where steeper power-law indexes up to $-9/4$ can be observed \citep{Coughlin_2019}. The goal of this work is to systematically select canonical TDE candidates following this predicted single flare behavior. eROSITA TDE candidates that show larger diversity in X-ray light curves (e.g., \citealt{guolo2024systematic}), will be discussed in a later publication.

The eROSITA light curves include fluxes of all four complete all-sky surveys (eRASS1-eRASS4) and eRASS5 if available (henceforth eRASS4(5)). This provides a coverage of 1.5--2.0\,years, with each eRASS being approximately six months after the previous one. Although the eROSITA sampling of light curves is too sparse to precisely constrain rise and decay times and slopes and only provides the lower limit for the peak flux, it allowed us to differentiate canonical TDE candidates from other sources that vary repeatedly or stay consistently bright on a longer timescale. We used the light-curve classification from the eRO-ExTra catalog, which divides the sample into four classes using eRASS1-4(5) data: decline, flare, brightening, and other. Canonical TDEs were selected from the flare and decline classes. The former class implies that eROSITA has covered the rise and decay phases of the event, and sources in the latter class were only observed from or after their peak. Here, we assumed the monotonic behavior based only on the eROSITA information, with the caveat that short-term variability can take place between eRASSes. This was seen before in other TDE candidates \citep{Liu_2023}. Similarly, we did not use the short term variability within a single eRASS to discriminate between sources.

\begin{table*}
\caption{Excerpt of X-ray properties of 31 TDE candidates selected from eRO-ExTra catalog. }
\begin{tabular}{llcccccc}
\hline
\hline

  \multicolumn{1}{c}{} &
  \multicolumn{1}{c}{Name} &
  \multicolumn{1}{c}{RA} &
  \multicolumn{1}{c}{Dec} &
  \multicolumn{1}{c}{$\mathrm{MJD_{peak}}$} &
  \multicolumn{1}{c}{$\mathrm{N_H}$} &
  \multicolumn{1}{c}{$\mathrm{\Gamma_{peak}}$} &
  \multicolumn{1}{c}{$\mathrm{F_{peak, 0.2-2.3 keV}}$} \\

  \multicolumn{1}{c}{} &
  \multicolumn{1}{c}{} &
  \multicolumn{1}{c}{[deg]} &
  \multicolumn{1}{c}{[deg]} &
  \multicolumn{1}{c}{[days]} &
  \multicolumn{1}{c}{[$\mathrm{10^{20} cm^{-2}}$]} &
  \multicolumn{1}{c}{} &
  \multicolumn{1}{c}{[$10^{-13}$ $\mathrm{erg \; cm^{-2}s^{-1}}$]} \\
\hline
  1 & eRASSU J011430.8-593654 & 18.628275 & -59.614870 & 59176.2 & 2.4 & 4.11 $\pm$ 0.29 & 6.37 $\pm$ 0.88\\
  2 & 1eRASS J014133.4-443413 & 25.389889 & -44.570661 & 59008.8 & 1.6 & 4.15 $\pm$ 0.46 & 3.11 $\pm$ 0.64\\
  3 & 1eRASS J024049.7-595432 & 40.205692 & -59.908882 & 59190.0 & 2.6 & 2.86 $\pm$ 0.30 & 2.78 $\pm$ 0.42\\
  4 & 1eRASS J031857.2-205459 & 49.739309 & -20.914231 & 59061.3 & 2.2 & 4.07 $\pm$ 0.26 & 4.18 $\pm$ 0.54\\
  5 & 1eRASS J034004.8-380004 & 55.020585 & -38.000508 & 59057.0 & 1.4 & 2.93 $\pm$ 0.18 & 4.50 $\pm$ 0.46\\
  6 & 1eRASS J034117.4-455718 & 55.321538 & -45.955209 & 59046.9 & 1.2 & 2.65 $\pm$ 0.39 & 1.15 $\pm$ 0.23\\
  7 & 1eRASS J043959.6-651403 & 69.997962 & -65.234243 & 58830.2 & 4.8 & 4.89 $\pm$ 0.08 & 13.17 $\pm$ 0.44\\
  8 & eRASSU J060829.4-435320 & 92.122520 & -43.888803 & 59130.4 & 9.0 & 4.09 $\pm$ 0.23 & 6.51 $\pm$ 0.61\\
  9 & eRASSU J061508.1-445921 & 93.783636 & -44.989211 & 59133.2 & 5.7 & 3.24 $\pm$ 0.38 & 1.97 $\pm$ 0.35\\
  10 & eRASSU J063413.3-713908 & 98.555526 & -71.652184 & 59094.8 & 10.1 & 4.25 $\pm$ 0.15 & 2.34 $\pm$ 0.14\\
  11 & 1eRASS J064449.4-603704 & 101.205323 & -60.617569 & 58974.3 & 11.7 & 6.13 $\pm$ 0.62 & 0.61 $\pm$ 0.12\\
  12 & eRASSU J070709.4-842044 & 106.789280 & -84.345486 & 59120.8 & 15.4 & 4.14 $\pm$ 0.26 & 2.44 $\pm$ 0.24\\
  13 & eRASSU J074426.3+291606 & 116.109686 & 29.268236 & 59143.7 & 4.3 & 4.25 $\pm$ 0.27 & 12.34 $\pm$ 1.49\\
  14 & eRASSU J082055.8+192538 & 125.232529 & 19.427145 & 59150.5 & 5.0 & 5.17 $\pm$ 0.33 & 10.66 $\pm$ 1.39\\
  15 & 1eRASS J082336.8+042303 & 125.903275 & 4.383791 & 59153.2 & 2.6 & 4.12 $\pm$ 0.09 & 100.83 $\pm$ 4.27\\
  16 & 1eRASS J084717.3-751050 & 131.827087 & -75.180849 & 59078.1 & 12.3 & 2.75 $\pm$ 0.46 & 0.69 $\pm$ 0.13\\
  17 & 1eRASS J091657.8+060955 & 139.240810 & 6.165557 & 58977.4 & 3.5 & 5.40 $\pm$ 0.40 & 8.89 $\pm$ 1.35\\
  18 & 1eRASS J110240.0-051809 & 165.666908 & -5.303827 & 58999.7 & 4.3 & 4.92 $\pm$ 0.44 & 5.27 $\pm$ 0.96\\
  19 & 1eRASS J130535.7+014700 & 196.399491 & 1.783504 & 59025.5 & 1.9 & 3.68 $\pm$ 0.25 & 7.71 $\pm$ 1.01\\
  20 & eRASSU J140158.1+081402 & 210.492210 & 8.233985 & 59038.6 & 2.1 & 5.19 $\pm$ 0.43 & 4.34 $\pm$ 0.72\\
  21 & 1eRASS J142140.3-295325 & 215.415121 & -29.889236 & 59066.4 & 7.1 & 3.10 $\pm$ 0.12 & 15.74 $\pm$ 0.95\\
  22 & 1eRASS J143623.9-174103 & 219.099536 & -17.684774 & 58881.1 & 8.2 & 4.55 $\pm$ 0.30 & 5.04 $\pm$ 0.61\\
  23 & eRASSU J145622.8-283853 & 224.095125 & -28.648065 & 59076.5 & 12.6 & 5.30 $\pm$ 0.30 & 6.30 $\pm$ 0.67\\
  24 & eRASSU J145954.5-260822 & 224.977157 & -26.139575 & 59076.5 & 14.9 & 2.94 $\pm$ 0.54 & 1.78 $\pm$ 0.39\\
  25 & 1eRASS J153406.2-090332 & 233.525738 & -9.059064 & 58897.8 & 12.5 & 4.24 $\pm$ 0.12 & 17.96 $\pm$ 0.87\\
  26 & eRASSU J164649.4-692539 & 251.705774 & -69.427495 & 59115.8 & 10.8 & 4.47 $\pm$ 0.19 & 10.47 $\pm$ 0.72\\
  27 & 1eRASS J173029.5-850501 & 262.607947 & -85.084201 & 59125.0 & 11.3 & 3.00 $\pm$ 0.40 & 1.93 $\pm$ 0.34\\
  28 & 1eRASS J190146.6-552200 & 285.444381 & -55.366377 & 58948.6 & 8.2 & 5.05 $\pm$ 0.29 & 10.24 $\pm$ 1.17\\
  29 & 1eRASS J192128.0-502746 & 290.367720 & -50.465458 & 59138.3 & 6.8 & 2.81 $\pm$ 0.25 & 9.09 $\pm$ 1.02\\
  30 & 1eRASS J210857.6-562835 & 317.243526 & -56.475421 & 59149.3 & 4.0 & 3.24 $\pm$ 0.28 & 6.66 $\pm$ 0.89\\
  31 & eRASSU J234403.1-352640 & 356.012960 & -35.444551 & 59180.4 & 1.2 & 4.62 $\pm$ 0.08 & 189.75 $\pm$ 7.21\\
\hline\end{tabular}

\tablefoot{The table includes eROSITA names with the X-ray detection coordinates (RA, Dec) from eRASS:5, a cumulative catalog of 4(5) eRASSes; the peak date ($\mathrm{MJD_{peak}}$); the best-fit photon index ($\Gamma$), and the flux in 0.2--2.3 keV at the peak eRASS ($\mathrm{F_{peak, 0.2-2.3 keV}}$). The additional properties of these sources can be found in the eRO-ExTra catalog.}
\label{tab:tde_xray}
\end{table*}

\subsection{eROSITA spectral modeling}
The X-ray spectra at peak luminosity made it possible for us to highlight the strongest TDE candidates from the eRO-ExTra catalog. All our sources were modeled with fixed Galactic-absorption (\texttt{tbabs}) components with equivalent neutral hydrogen column densities ($\mathrm{N_{H,gal}}$) extracted from the HI4PI survey \citep{HI4PICollaboration} at their position. The eRO-ExTra catalog provides best-fit values with a power-law model \texttt{tbabs*powerlaw}. To identify canonical X-ray soft TDEs, we selected sources with photon indexes $\Gamma>3$ taking into account  $1\sigma$ errors. This threshold was chosen based on the typical observational properties of previously X-ray-selected TDEs ($\Gamma=3-5$; \citealt{ 2020SSRv..216...85S}). We note that this threshold again selects only the canonical X-ray soft population of TDEs and not the entire class of possible X-ray TDEs, as several evolution scenarios may cause spectra to be harder. For example, hard spectra can arise from the Comptonization of disk photons in a newly-formed corona \citep{guolo2024systematic} or in jetted TDEs, which have harder spectra with $\Gamma<2$ even at the light-curve peak \citep{2011Natur.476..421B,Saxton_2012,Bradley_Cenko_2012}.

In addition to the peak eROSITA epochs, we also modeled the spectra of all other eROSITA detections with sufficient counts (\texttt{DET\_LIKE}>15). For all epochs, we extracted spectra using the \texttt{eSASS} \citep{2022A&A...661A...1B} task \texttt{SRCTOOL} (version eSASSusers\_211214) with circular source and annular background regions centered on the X-ray positions. A redshifted black-body model, \texttt{tbabs*zbbody,} was used for all fits. Utilizing \texttt{SRCTOOL}'s AUTO mode, the radii of the source and background regions were selected taking into account the source counts, the level of the background map model at the source position, the best fitting source extent model radius, and a model of the effective eROSITA PSF at the source position (averaged over the exposure, and summed over all operating telescope modules). The spectra were fit unbinned and using the Bayesian X-ray analysis software \citep[BXA;][]{2014A&A...564A.125B}, which connects the nested sampling algorithm UltraNest \citep{buchner2021ultranest} with the fitting environment XSPEC \citep{1996ASPC..101...17A}. C-statistic \citep{1976cash} was used, and the eROSITA background was modeled using the principal component analysis (PCA) technique described in \citet{2018A&A...618A..66S}. Each extracted eROSITA spectrum, which contains a contribution from both the source and background emission, was jointly fit using the source and background models. We assumed wide flat priors on the photon index and the logarithm of the normalization. The best fitting parameters are presented in Table~\ref{tab:modeling_xray}.

\subsection{Host photometry}

 In addition to the LS10 values of W1-W2 provided in eRO-ExTra, we explored individual data points in available mid-IR light curves (see Sect.~\ref{sec:mid_ir}) to confirm that the pre-TDE value of W1-W2 is less than 0.8 and to exclude possibly missed AGNs varying in the mid-IR. Finally, one source was excluded since its potential optical counterpart is very faint and inconsistent with a canonical TDE host in a low-redshift universe.

\section{Final sample of golden TDE candidates}
\label{sec:final_sample}
The sample of canonical TDE candidates selected from eRO-ExTra includes 31 sources that 1) have a single-decline or a single-flare eROSITA light curve; 2) have $\Gamma>3$ accounting for errors; 3) do not have significant archival X-ray detections at the eROSITA peak level; and 4) have a clear host-galaxy counterpart without known AGN signatures. The sample and its essential X-ray properties from the eRO-ExTra catalog are presented in Table~\ref{tab:tde_xray}.

\subsection{Optical counterparts for the TDE sample}
Each TDE candidate is associated with an LS10 optical counterpart. In agreement with the reliability criterion (\texttt{p\_any}>0.17) used in \citet{Grotova_2025}, 28 of 31 sources have a reliable counterpart. However, it is important to note that three optical counterparts\footnote{See Appendix~\ref{sec:j1421} for more details on the optical counterpart of J1421.} of sources with low \texttt{p\_any}<0.17 are still likely genuine, since the optical counterpart position is in agreement with the X-ray position within $3\sigma$. The low \texttt{p\_any} indicates that the multiwavelength properties of the counterparts do not entirely match the properties of typical of X-ray emitters. This is not surprising, as no TDEs were included in the training sample used in \citet{Salvato2018}. LS10 image cutouts of the host galaxies are shown in Fig.~\ref{fig:ls10_cutouts}, and the counterpart properties are listed in Table~\ref{tab:tde_ctps}. 

\begin{table}
\caption{LS10 counterparts of TDE candidates with their probabilities (\texttt{p\_any}) and redshifts (z).}
\begin{tabular}{llrrrr}
\hline
\hline
\multicolumn{1}{c}{} &
  \multicolumn{1}{c}{Name} &
  \multicolumn{1}{c}{\texttt{p\_any}} &
  \multicolumn{1}{c}{LS10\_RA} &
  \multicolumn{1}{c}{LS10\_Dec} &
  \multicolumn{1}{c}{z} \\

  \multicolumn{1}{c}{} &
  \multicolumn{1}{c}{} &
  \multicolumn{1}{c}{} &
  \multicolumn{1}{c}{} &
  \multicolumn{1}{c}{[deg]} &
  \multicolumn{1}{c}{[deg]} 
 \\
\hline

  1&J0114& 0.43 & 18.6286 & -59.6152 & 0.16\\
  2&J0141& 0.38 & 25.3891 & -44.5707 & 0.09\\
  3&J0240& 0.99 & 40.2062 & -59.9091 & $0.21^{+0.07}_{-0.04}$\\
  4&J0318& 0.32 & 49.7388 & -20.9143 & $0.27^{+0.13}_{-0.09}$\\
  5&J0340 & 0.99 & 55.0208 & -38.0005 & $0.21^{+0.06}_{-0.05}$\\
  6&J0341& 0.99 & 55.3219 & -45.9553 & $0.34^{+0.08}_{-0.11}$\\
  7&J0439& 0.98 & 69.9986 & -65.2342 & 0.15\\
  8&J0608& 0.77 & 92.1226 & -43.8892 & 0.08\\
  9&J0615& 0.92 & 93.7837 & -44.9893 & $0.10^{+0.03}_{-0.03}$\\
  10&J0634 & 0.70 & 98.5558 & -71.6518 & 0.13\\
  11&J0644 & 0.58 & 101.2052 & -60.6178 & 0.12\\
  12&J0707& 0.24 & 106.7895 & -84.346 & $0.17^{+0.23}_{-0.02}$\\
  13&J0744 & - & 116.1089 & 29.2688 & 0.04\\
  14&J0820 & 0.95 & 125.2324 & 19.427 & 0.12\\
  15&J0823 & 0.99 & 125.9032 & 4.384 & 0.03\\
  16&J0847& 0.99 & 131.8265 & -75.1814 & -\\
  17&J0916& 0.16 & 139.2407 & 6.1658 & 0.09\\
  18&J1102& 0.26 & 165.6673 & -5.3031 & $0.33^{+0.10}_{-0.09}$\\
  19&J1305& 0.06 & 196.3994 & 1.7837 & $0.33^{+0.07}_{-0.08}$\\

  20&J1401 & 0.96 & 210.492 & 8.2337 & $0.26^{+0.06}_{-0.05}$\\
  21&J1421& 0.12 & 215.4155 & -29.8896 & 0.06\\
  22&J1436& 0.85 & 219.0989 & -17.6845 & $0.19^{+0.06}_{-0.03}$\\
  23&J1456& 0.95 & 224.0958 & -28.6487 & 0.09\\
  24&J1459& 0.02 & 224.978 & -26.1409 & 0.09\\
  25&J1534& 0.99 & 233.5261 & -9.0593 & 0.02\\
  26&J1646 & 0.21 & 251.7032 & -69.4271 & 0.02\\
  27&J1730 & 0.88 & 262.6191 & -85.0844 & $0.33^{+0.07}_{-0.05}$\\
  28&J1901 & 0.9 & 285.4461 & -55.3662 & 0.06\\
  29&J1921& 0.97 & 290.3667 & -50.4652 & 0.06\\
  30&J2108 & 0.98 & 317.2424 & -56.4753 & 0.04\\
  31&J2344 & 0.49 & 356.0123 & -35.4449 & 0.10\\
\hline\end{tabular}
\tablefoot{Photometric redshifts are noted with their $1\sigma$ errors, others are spectroscopic. The origin of redshifts as well as more counterpart properties can be found in the eRO-ExTra catalog. The photo-z for J0847 is not-reliable, since two optical photometry {\it griz} points were missing in the calculation; it is therefore not listed. The \texttt{p\_any} value for J0744 is omitted, since the counterpart was identified in \citet{Malyali_2023}.}
\label{tab:tde_ctps}
\end{table}

\begin{figure}
\centering
    \includegraphics[width=\linewidth]{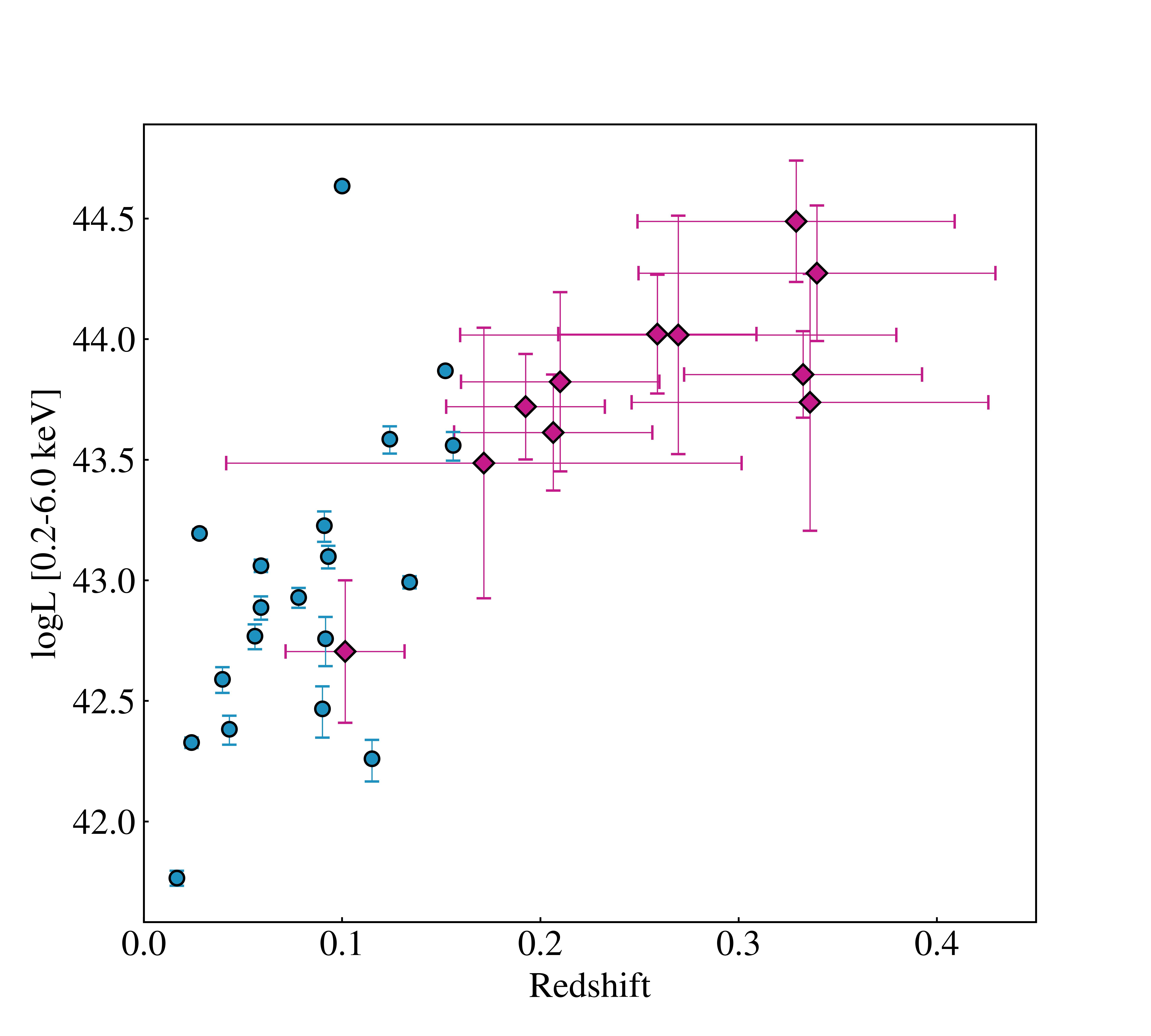}  
\caption{Rest-frame luminosity in 0.2-6.0 keV versus redshift. The blue circles denote sources with spectroscopic redshifts and the pink diamonds show those with photometric redshifts. The most luminous source in the sample is J2344.}
\label{fig:luminosity_z}
\end{figure}

Spectroscopic redshifts are available for 19 sources from either archival data or a dedicated follow-up program (see the summary provided in \citealt{Grotova_2025}); photo-z (Salvato et al., in prep.) were computed for the remaining sample using LS10 and Wide-field Infrared Survey Explorer (WISE; \citealt{2010AJ....140.1868W}) photometry with CIRCLEZ \citep{2024A&A...690A.365S} on LS10 photometry ( {\it griz}, W1,W2,W3,W4). Figure~\ref{fig:luminosity_z} shows the sample in the redshift-luminosity plane. Overall, the TDE sample spreads over the 0.02 < z < 0.34 redshift range and luminosities of $\mathrm{5.7\times10^{41}<L_{X}<5.3\times10^{44} ergs^{-1}}$ in the $0.2 - 6.0$ keV rest frame.

Several sources from the sample were previously reported in individual papers: 1eRASS J082336.8+042303 (AT2019avd; \citealt{Malyali_AT2019avd}), eRASSt J074426.3+291606 \citep{Malyali_2023}, and  eRASSt J234403.1-352640 \citep{homan2344, Goodwin_2023}. The TDE nature of the candidates was also strengthened by the presence of TDE-like emission in optical, mid-IR, and radio bands (more details in  Sect.\ref{sec:multivar}). This demonstrates the reliability of the methodology used in the presented selection.

\subsection{X-ray properties of the TDE sample}
\label{sec:xray_props}

This subsection provides insights into the X-ray properties of the selected TDEs, with a particular focus on light-curve shapes, rise and decay timescales, and spectral evolution.
Most sources in the sample do not show significant spectral variability, with their black-body temperature staying consistent within $3\sigma$ in all detected epochs (see Table~\ref{tab:modeling_xray}). Therefore, the vast majority of selected TDEs remain X-ray soft throughout the event. Only two candidates, namely J2344 and J0439, show kT evolution with $6\sigma$ (from eRASS1 to eRASS2) and $9\sigma$ (from eRASS2 to eRASS3), respectively. Whereas J2344 shows even further spectral softening from 80 eV to 40 eV within a year (from eRASS2 to eRASS4), J0439 exhibits a significant temperature increase from 80 eV to 210 eV. The observed spectral hardening might be associated with the formation of a hot corona (e.g., \citealt{Wevers_2021,Liu_2023}), similar to a transition from a soft, disk-dominated state to a hard, corona-dominated state observed in X-ray binary systems with a stellar-mass BH \citep{Wang_2022}.

X-ray light curves based on the eRASS4(5) eROSITA all-sky surveys in the soft 0.2–2.3 keV energy band with six-month cadence are provided in Fig.~\ref{fig:xray_lc}. By selection, all sources in the sample peaked either in eRASS1 (P1) or eRASS2 (P2). The P1 candidates show a decaying light curve, and the P2 candidates have a flare light curve. After the peak, the sources declined monotonically or declined after a  $0.5-1$ year plateau phase. It is important to note that the peak flux and amplitudes of the events are lower limits only, since the actual peak could happen between eROSITA scans.

The sample contains more P2 sources (23 objects) than P1 sources (8). The bias against the P1 group appeared from the selection criteria of the eRO-ExTra parent catalog; this includes, namely, amplitude (A>4) and significance (S>4) cuts between eRASS1 and eRASS2. Whereas P2 sources were selected based on their rise between eRASS1 and eRASS2, P1 sources were chosen based on their decay between eRASS1 and eRASS2. We find that the rise timescales for the sample (only possible to estimate for P2) is, on average, shorter than the decay timescales. Namely, all P2 sources rise within $\approx6$ months. However, the decay time for a fraction of TDEs is longer, with 30\,\% of P2 sources decaying at least over $1-1.5$ years. To confirm the bias, we tested whether P2 sources would be selected based on their decline amplitude: only nine of 24 sources would be kept after applying the A>4 and S>4 cuts in the decay phase between eRASS2 and eRASS3,  which is consistent with the number of P1 sources.

As mentioned before, the estimation of rise, decay, and plateau timescales is limited by the sparsity of the eROSITA light-curve sampling. Therefore,  further timescales are only provided to assess general trends; possible variability between eRASSes is thus omitted. For the P2 sources, the rising timescale does not exceed six months (or 1 year if assuming that the peak can also be between eRASS2 and eRASS3). This aligns with the expected rise time for TDEs \citep{Gezari_2021}. Three P2 sources show a plateau phase at the peak level lasting $\approx 0.5-1$ year. The majority of sources (21 of 31) decay within half a year, which is in agreement with the TDE light-curve trends from \citet{Sazonov_2021}. The remaining sources decay over two to three eRASSes ($1-1.5$ years).

The majority of pre-eROSITA archival X-ray data for the sample provided in the eRO-ExTra catalog\footnote{No data are available for J0744+29: the archival X-ray detection belongs to the bright galaxy nearby, while the real host is an X-ray quiet dwarf galaxy \citep{Malyali_2023}.
} are not constraining and only include ULs that are either shallower than the brightest eROSITA detection or significantly above eROSITA's detected minimum. There are only constraining measurements for four sources in the sample. An archival UL by {\it XMM-Newton} revealed a large amplitude of more than a factor of 50 of the flare in 1eRASS J082336.8+042303, whereas the amplitude between eRASS1 and eRASS2 was of only a factor of six. Deep archival {\it ROSAT} ULs of 1eRASS J091657.8+060955 and eRASSU J164649.4-692539 reinforce the archival X-ray quiescence of the candidates. Moreover, in the case of  1eRASS J091657.8+060955, this allowed us to prove the presence of the flare, since eROSITA has only observed the decline phase. In addition, 1eRASS J024049.7-595432 has a constraining {\it XMM-Newton} detection. The flux is comparable to the detected eROSITA minimum and can be interpreted as an underlying host emission being consistent with the luminosity expected from star-forming galaxies \citep{10.1093/mnras/stu1999}. The archival constraints are shown in the light-curve plots in Fig.~\ref{fig:xray_lc}.

\section{Multiwavelength variability}
\label{sec:multivar}

In this section, we describe the optical, mid-IR, and radio properties of the sample.

\subsection{Mid-IR variability}
\label{sec:mid_ir}

Mid-IR flares in TDEs are thought to arise from the reprocession of optical, ultraviolet (UV), and X-ray photons by dust within a few parsecs from the SMBH. They can last for years \citep{2016ApJ...828L..14J,Dou_2017,2021SSRv..217...63V}, peak at $3-10 \mathrm{\mu m}$, and reach luminosities of $10^{42-43}$ $\mathrm{erg\,s^{-1}}$ \citep{lu2016}.

To study the mid-IR properties of the eROSITA TDE sample, we used data from WISE during the NEOWISE reactivation mission (NEOWISE-R; \citealt{neowise}), which covered the entire sky every six months from Sept 2014 to Aug 2024 in the W1 (3.4 $\mu$m) and W2 (4.6 $\mu$m) bands. Light curves were extracted using the NASA/IPAC Infrared Science Archive within 5" of the position of optical counterparts. The photometry was rebinned to one data point per NEOWISE-R all-sky scan using a weighted mean. The W1 and W2 light curves for all sources are shown in Figs.~\ref{fig:wise_lc1} and \ref{fig:wise_lc2}.

\begin{figure}
\centering
    \includegraphics[width=\linewidth]{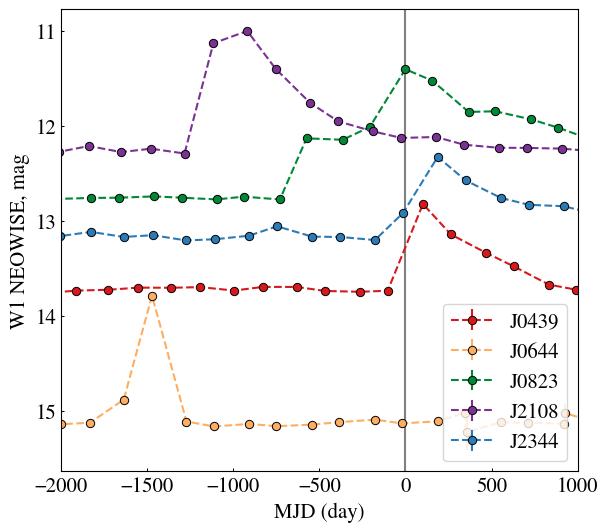}  
\caption{W1 NEOWISE light curves of five TDE candidates with significant IR flares of more than 0.9 mag. The X-axis is shifted so that x=0 corresponds to the peak eROSITA detection for each source. The magnitudes are given in the Vega system.}
\label{fig:wise_flares}
\end{figure}

Approximately $30\,\%$ of the eROSITA TDE sample shows mid-IR variability. Sources with the significance $\mathrm{\sigma_{W1,max}>3}$ and amplitude $\mathrm{\Delta W1>0.3}$mag are summarized in Table~\ref{tab:wise_flares}. For five of those, we found flares with $\mathrm{\Delta W1\approx0.9-1.5}$ mag (see Fig.~\ref{fig:wise_flares}). Most mid-IR flares show a fast rise within six months, followed by a slow decay over several years. The exception is J0644, with its brightening lasting about one year and a fading time of approximately six months. The "fast rise and slow decay" trend resembles the one found for the mid-IR selected TDEs and is consistent with dust reprocessing of the intrinsic TDE flare \citep{masterson2024}. 

\begin{table*}
\centering
\caption{Properties of prominent mid-IR flares of TDE candidates.}
\begin{tabular}{cccccc}
\hline
\hline
 Name & $\mathrm{\Delta_{W1}}$ [mag$\mathrm{_{Vega}}$] & $\mathrm{\sigma_{W1}}$ & $\mathrm{MJD_{W1,peak}}$ & $\mathrm{W1-W2_{pre}}$ & $\mathrm{W1-W2_{flare}}$ \\
\hline
J0439 & 1.0 & 99.4 & 59021 & 0.4 & 0.7 \\
J0644 & 1.5 & 58.3  & 57499 & - & - \\
J0823 & 1.4 & 78.0  & 59151 & 0.1 & 0.8 \\
J2108 & 1.3 & 152.4 & 58229 & 0.1 & 0.7 \\
J2344 & 0.9 & 79.0 & 59365 & 0.3 & 0.7 \\
\hline
J0141 & 0.4 & 3.8 & 59184 & - & - \\
J0608 & 0.3 & 17.2  & 59276 & 0.3 & 0.5 \\
J0615 & 0.6 & 38.0 & 59121 & 0.2 & 0.6 \\
J0847 & 0.4 & 6.0  & 59255 & - & - \\
J1436 & 0.3 & 9.2 & 59250 & 0.3 & 0.6 \\
\hline
\end{tabular}

\label{tab:wise_flares}
\end{table*}

Seven sources with mid-IR flares have well-sampled light curves in both the W1 and W2 bands that can be used to explore the W1-W2 evolution; these can trace the development of temporary accretion-driven activity during the TDE accretion phase. For all seven sources, the mid-IR brightening was accompanied by a significant reddening (see Table~\ref{tab:wise_flares}), with the most significant changes in W1-W2 observed for J0823 (from 0.1\,mag to 0.8\,mag) and J2108 (from 0.1\,mag to 0.7\,mag).

\subsection{Optical variability}
\label{sec:opt_flares}

Next, we explored how many sources in our sample of TDEs are accompanied by optical transient emission using data from Gaia \citep{gaia_2016}, the Asteroid Terrestrial Impact Last Alert System (ATLAS; \citealt{2018PASP..130f4505T}), and the \textit{Zwicky} Transient Facility (ZTF; \citealt{2019PASP..131a8002B}). ZTF host-subtracted light curves in the {\it g} band were extracted at the position of the optical counterpart using the ZTF forced photometry service \citep{2019PASP..131a8003M}. We applied a signal-to-noise threshold of $\mathrm{S/N > 3}$, $\mathrm{scisigpix<10}$, $\mathrm{zpmaginpscirms<0.1}$ to filter out noisy images (more information on the data quality parameters is provided in the ZTF manual\footnote{\url{https://irsa.ipac.caltech.edu/data/ZTF/docs/forcedphot.pdf}}).  Gaia {\it g}-band light curves were extracted from the Gaia Alerts web server\footnote{\url{http://gsaweb.ast.cam.ac.uk/alerts/home}}. Since the service does not provide host-subtracted light curves, we estimated the flare magnitudes by subtracting the provided pre-flare historical magnitudes of the hosts. ATLAS light curves were obtained from publicly available photometric data in the cyan (420--650 nm) and orange (560--820 nm) filters. They were extracted by forced photometry on the difference images and rebinned to 24\,hr. To remove epochs of low-quality photometry, we discarded data points with the semi-major axis of the fit PSF model that were greater than three pixels. 

\begin{figure}
\centering
    \includegraphics[width=\linewidth]{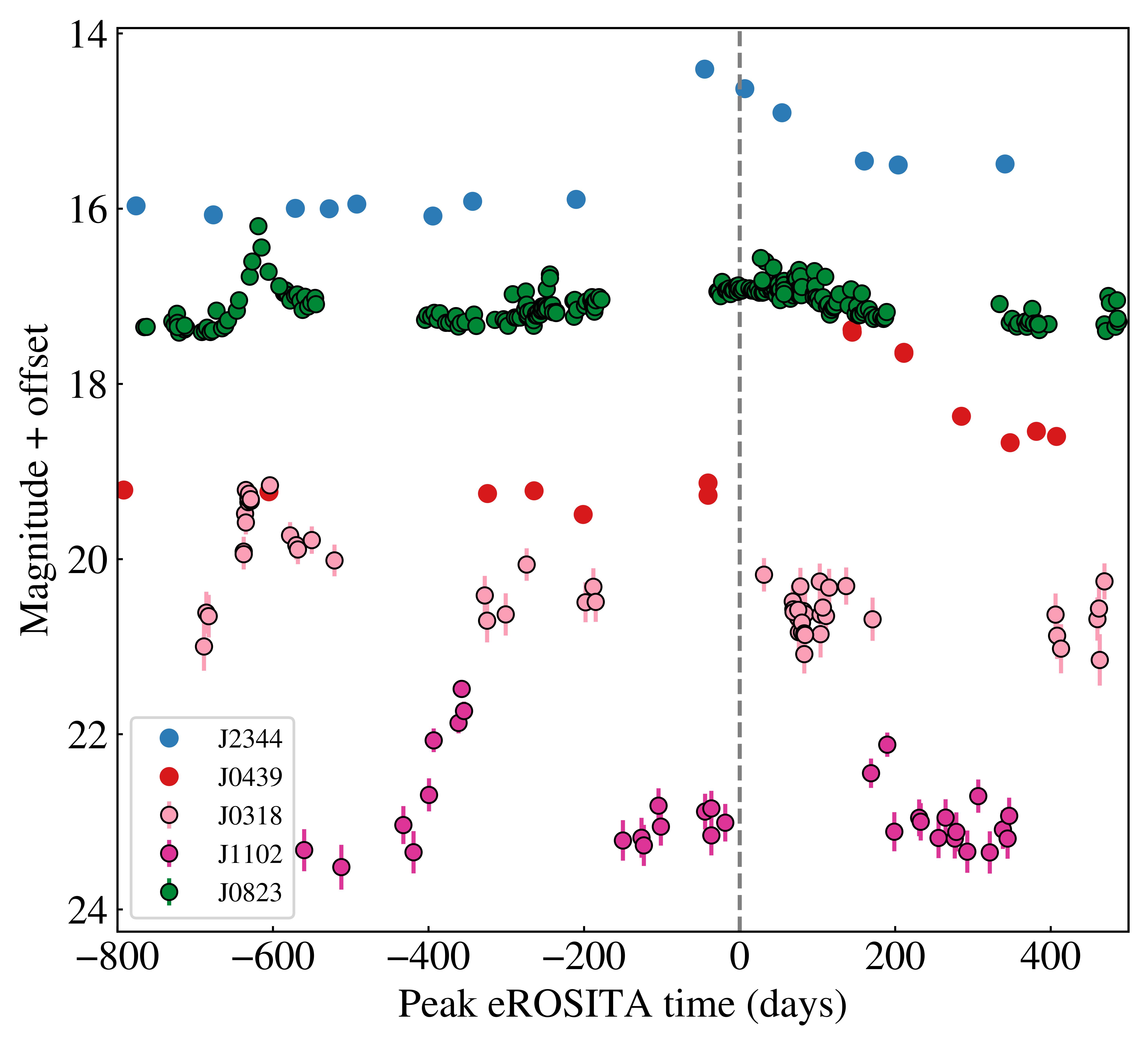}  
\caption{Non-host-subtracted g-band light curves of five TDE candidates showing significant optical flares. The x-axis shows time after eROSITA peak in days, so x=0 corresponds to the peak eROSITA detection for each source. The summary of observational information is provided in Table~\ref{tab:optical_lc_ul}. The individual-host-subtracted light curves are showin in Fig.~\ref{fig:optical_lc_ztf_gaia}. The optical flare of J0744 is too faint to be distinguished in a non-host-subtracted light curve and is only shown in the host-subtracted plot. }
\label{fig:optical_flares}
\end{figure}

Flares with flux amplitudes larger than two were found in the ZTF and Gaia data for six sources (see the summary in Table~\ref{tab:optical_flares}). Their single-band, non-host-subtracted light curves are compiled in Fig.~\ref{fig:optical_flares}, while more detailed plots of host-subtracted light curves are available in Fig.~\ref{fig:optical_lc_ztf_gaia}. Detailed discussions of the optical properties of J0744, J0823, and J2344 have already been presented elsewhere \citep{Malyali_2023, Malyali_AT2019avd,homan2344}.  

The majority of prominent flares discussed above are also clearly detected with ATLAS. Although ATLAS provides coverage for all sources not covered by Gaia or ZTF, the remaining TDE candidates either did not have accompanying optical flares, or ATLAS data are too noisy to identify additional fainter and statistically significant flares. For these sources, we estimated ATLAS ULs for potentially undetected flares using a mean, five-sigma limit magnitude in the orange filter. The UL values are presented in Table~\ref{tab:optical_lc_ul} and span from 18.2 to 19.1 mag. The shallow level of ATLAS ULs in comparison to, for example, the ZTF typical limiting magnitude of $\approx 20.5-21$ in the {\it g,r} bands, shows a currently existing observational bias against detecting fainter optical transients in the eROSITA\_DE sky. Given the variety of the timing of optical and X-ray flares (from around -600 to 0 days; see Fig.~\ref{fig:optical_flares}), all-sky deeper optical surveys would be required to systematically identify optical flares for X-ray-selected TDEs. As an example of an observational bias, the LS10 optical transient J1421 was detected in the {\it g} band with 19.21 mag due to the accidentally fortunate timing of LS10 epochs (see Appendix~\ref{sec:j1421}) and would have been missed had we used an ATLAS UL equal to 18.3 mag. The transient is off-center and is a strong candidate for an off-nucleus TDE (more details in Sect.~\ref{sec:j1421_imbh}).

Finally, we estimated $\mathrm{L_{opt,peak}/L_{x,peak}}$ at the optical-flare peak time for the quasi-simultaneous X-ray and optical flares (J0439, J0744, J2344). The value ranges in $\mathrm{L_{opt,peak}/L_{X,peak} \in (4-11)}$. This is in the agreement with the $\mathrm{L_{opt}/L_{x}}$ estimates from \citet{guolo2024systematic}. The upper limits for $\mathrm{L_{opt,peak}/L_{X,peak}}$ were calculated for the remaining sample. The resulting values, ranging between five and 140, are consistent with the expectations. This confirms that although optical flares were not detected for the majority of the sample, this might be due to observational limitations.

\begin{table}
\centering
\caption{Properties at peak of optical flares of eROSITA TDE candidates.}
\begin{tabular}{lcrr}
\hline
\hline
  \multicolumn{1}{c}{Name} &
  \multicolumn{1}{c}{{\it g}-band} &
  \multicolumn{1}{c}{$\mathrm{L_{g, peak}}$} &
  \multicolumn{1}{c}{$\mathrm{Date_{peak}}$} 
  \\
  \multicolumn{1}{c}{} &
  \multicolumn{1}{c}{[mag$_{\rm AB}$]} &
  \multicolumn{1}{c}{$\mathrm{\times 10^{44}[erg s^{-1}}$]} &
  \multicolumn{1}{c}{[MJD]} 
  \\
  \hline
  J0318 & $20.3 \pm 0.1$ & $5.0_{-3.6}^{+7.1}$ & 58430 \\
  J0439 & $18.41\pm0.12$ & $7.97_{-0.83}^{+0.93}$ & 58975 \\
  J0744 & $19.6\pm 0.1$  & $0.16_{-0.01}^{+0.01}$ & 59126 \\
  J0823 &  $17.13\pm 0.04$& $0.86_{-0.03}^{+0.03}$ & 58538 \\
  J1102 & $19.8\pm 0.3$ & $13.54_{-8.48}^{+17.46}$ & 58619 \\
  J2344 & $16.50 \pm 0.11$& $18.74_{-1.80}^{+1.99}$ & 59135 \\

  \hline
  J1421 & $19.21 \pm 0.05$  & $0.53_{-0.03}^{+0.02}$ &58643 \\
  \hline
\end{tabular}

\label{tab:optical_flares}

\end{table}

\subsection{Radio variability}
To search for potential radio counterparts, the LS10 positions of the eROSITA TDEs were cross-matched with the Rapid ASKAP Continuum Survey \citep[RACS,][]{RACs1,RACs2,RACs3} source catalog with a radius of 25”. We chose this cross-matching radius as it represents the resolution of the RACS images (the full width at half maximum of the PSF is 25"; \citealt{RACs1}). Four radio counterparts out of the 31 TDE candidates were found. For these four sources, we then searched the CASDA Data Portal for all publicly available RACS observations covering their sky locations. Flux densities in the image plane were derived by fitting an elliptical Gaussian with  the size of the synthesized beam. The results are reported in Table \ref{tab:RACs}.

Each of the four detected sources has between two and four RACS observations at central observing frequencies of 0.9\ GHz or 1.4\ GHz, allowing us to probe for any radio variability. The TDE candidates J1646, J2108, and J1305 show indications of radio variability across the RACS observations spanning from 2019--2024. J1646 was already detected at 0.9\,GHz 539\,d before the eROSITA discovery, suggesting emission from the host unrelated to the TDE. This is consistent with the corresponding luminosity of L$\mathrm{_{\rm 9 GHz}=1.7\times10^{28}\,erg s^{-1}\,Hz^{-1}}$. Variability at 1.4\,GHz was detected between 62 and 82\,d after the eROSITA peak, in agreement with expectations from a newly launched outflow.  J2108 showed constant (within the errors) radio emission before and after the eROSITA peak date at a luminosity of L$\mathrm{_{\rm 9 GHz}=5.1\times10^{30}\,erg s^{-1}\,Hz^{-1}}$. In this case, the radio emission is also very likely unrelated to the X-ray flare. The centroid of the radio source associated with  J1305 is 11” away from the optical position of the galaxy the event is associated with, so this radio source may be unrelated to the host galaxy. Due to the resolution of the RACS images and the fact that this radio source falls within one beam of the optical counterpart, higher resolution radio observations are required to determine if this radio source is associated with the optical counterpart. For the fourth source, J0823, only RACS observations in different bands exist. However, we note that \citet{Wang2023} reported the detection of radio variability in this source from the VLA and VLBA interpreted as evidence for a radio-emitting outflow launched by the transient event. The remaining 27 TDE candidates in the sample were undetected by RACS, with typical 3$\mathrm{\sigma}$ upper limits of $\approx$500--1000\,$\mathrm{\mu}$Jy. A detailed analysis of the radio properties of these events, including more sensitive dedicated radio follow-up observations of 22 TDE candidates in this sample, is presented in \citet{goodwin2025}. In addition, a summary of the radio flare properties of J2344 was presented in \citet{Goodwin_2023}.

\begin{table}
\caption{RACS radio detections of eROSITA TDEs.}
\centering
\begin{tabular}{lrrrr}
\hline
\hline
Name & MJD & Frequency& Flux Density\\
 & [day] & [GHz] & [$\mu$Jy]\\
\hline
J0823 &  59232 & 1.4& 862$\pm$221\\
& 60310 & 0.9& 1335$\pm$277\\
\hline
J1305 & 58970 & 0.9 & 4280$\pm$241\\
& 60326 & 0.9 & 4110$\pm$332\\
\hline
J1646 &  58935 & 0.9& 1888$\pm$162\\
&  59211 & 1.4& 1221$\pm$160\\
&  59231 & 1.4& 2222$\pm$860\\
\hline
J2108 &  58610& 0.9& 1330$\pm$220\\
&  58936 & 0.9& 989$\pm$251\\
 & 59220 & 1.4& 952$\pm$167\\
 & 60330 & 0.9& 1170$\pm$150\\
\hline

\end{tabular}
\label{tab:RACs}
\end{table}

\section{Host-galaxy properties}
\label{sec:sed}

\subsection{SED modeling}

To describe the host galaxy properties of our TDE candidates, we modeled the spectral energy distributions (SEDs) using the photometry provided by LS10. This includes optical photometry in {\it griz} bands from DECam and mid-IR photometry W1-4 from NEOWISE. Fluxes were corrected for the Galactic foreground extinction. In addition to the inverse flux variance FLUX\_IVAR provided in the LS10 catalog, we introduced systematic uncertainties equal to 0.05\,mag for {\it griz} filters and 0.2\,mag for W1-4 filters \citep{Salvato_2022}, which account for factors such as source variability and different extraction radii. The modeling was not performed for J0847, J1646, and J1421 since two or more optical photometry bands are missing.

We applied the same methodology and models as used in several previous systematic TDE studies \citep{Sazonov_2021,vanVelzen_2021,Hammerstein_2023,yao2023,masterson2024}. We fit models from the flexible stellar population synthesis (FSPS) framework \citep{2010ApJ...712..833C} with the Prospector software \citep{prospector}. We assumed a delayed, exponentially declining star formation history with a Chabrier initial mass function \citep{2003PASP..115..763C}. The model has five free parameters: the galaxy's stellar mass, the age of the stellar population, the e-folding timescale of star formation, metallicity, and dust optical depth, for which we used the extinction law from \citet{2000ApJ...533..682C}. To derive uncertainties, we applied MCMC sampling using the emcee package \citep{2013PASP..125..306F}. The MCMC chains were initialized with 100 walkers, starting from the maximum likelihood fit with a burn-in phase of 500 steps per walker and 1000 iterations of the MCMC sampling.

We used MCMC fits to evaluate the host stellar masses and rest frame $\mathrm{^{0.0}}u-r$ colors, where 0.0 denotes the rest-frame values. As in previous studies, we reported a surviving stellar mass obtained by multiplying the total mass by the factor "mfrac" obtained from the model. The surviving mass is defined as a mass that survived throughout the galaxy's evolution, accounting for mass loss due to stellar winds or supernovae. The SED fitting results are shown in Table~\ref{tab:sed_results}.

\begin{figure}
\centering
    \includegraphics[width=\linewidth]{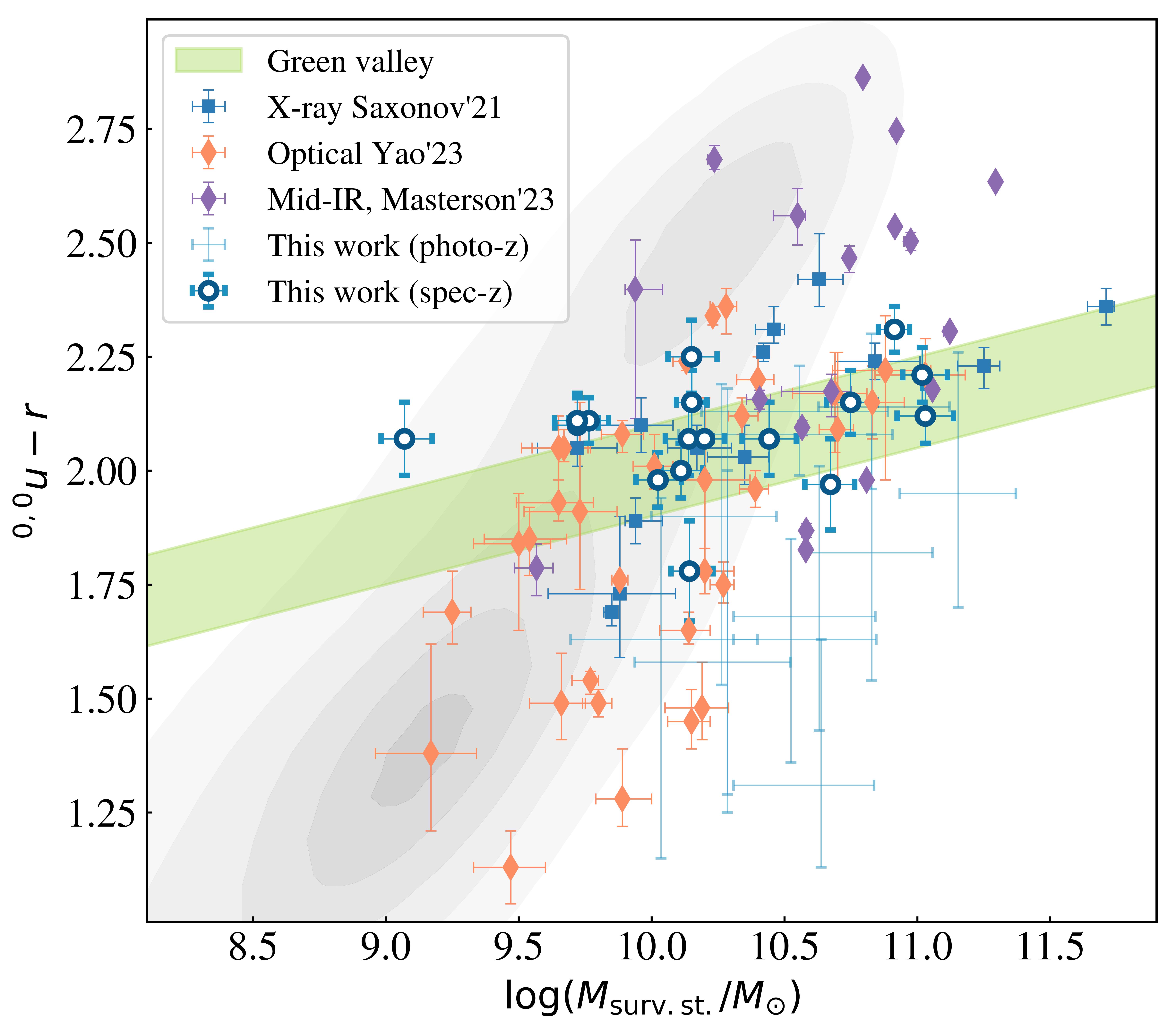}  
\caption{Host-galaxy color versus surviving stellar mass of the eROSITA TDE sample in comparison to other TDE samples selected in X-rays, the optical, and the mid-IR. Gray contours denote a sample of SDSS galaxies within the same redshift range (z<0.34) from \citet{2014ApJS..210....3M}. The green area shows the green valley defined in \citet{yao2023}. The blue circles show the eROSITA TDEs presented in this work with spec-z, and sources with photo-z are shown with blue error bars. The errors for sources with photo-z are calculated accounting for $1\sigma$ photo-z errors. The orange and purple diamonds and blue squares denote TDE samples selected in the optical \citep{yao2023} and mid-IR \citep{masterson2024} and in X-rays \citep{Sazonov_2021}, respectively.}
\label{fig:greenvalley}
\end{figure}

The comparison of the stellar masses and colors of this TDE sample with other X-ray-, optically, or mid-IR-selected TDE samples is shown in Fig.~\ref{fig:greenvalley}. The eROSITA TDE sample shows an overdensity in the green valley, with more than half of the modeled sample (18 of 28 sources) falling in this conservatively defined area. Previous studies also found that X-ray- and optically selected TDE populations are preferentially located in the green valley (e.g., \citealt{2014ApJ...793...38A, vanVelzen_2021,Sazonov_2021,yao2023,Hammerstein_2023}). The higher TDE occurrence rate in the green-valley galaxies is considered to be due to the higher concentration of stars scattered into the loss cone because of a recent merger \citep{2016ApJ...830L..32P, Hammerstein_2021} or a starburst \citep{2016ApJ...818L..21F,2018MNRAS.480.5060S,2016ApJ...825L..14S}. However, it was recently shown that higher central stellar concentration increases the rate of strong scattering that may lead to the decrease of the TDE rate scattered into the loss cone \citep{2024arXiv241105086T}. While X-ray- and optically selected TDEs are concentrated in the green valley, the mid-IR TDE population consists of redder and more massive galaxies, which was explained by the selection bias of choosing the brightest IR flares \citep{masterson2024}. However, X-ray TDEs with mid-IR flares in this work do not follow this trend, with their stellar masses being homogeneously distributed within $\mathrm{log(M_{gal}/M_{sun})\approx 9.6 - 11}$.

We note that the selection of the extinction model influences the stellar mass and color results. Thus, using the extinction law from \citet{1984A&A...132..389P}, the stellar masses reduce by 50\,\% on average, and $^{0,0}u-r$ increases by $\approx 0.1$\,mag. This results in a small movement toward the upper left corner of the diagram; however, the qualitative result of an overdensity in the green valley would not be changed. In addition, it is worth pointing out that the training sample used for computing the photo-z via machine learning was tuned for AGNs \citep{2024A&A...690A.365S}. A direct comparison between the photo-z and the spec-z for the sources with spectroscopic redshifts available suggests that the photo-z tends to be higher than should be, resulting in bluer colors. The increasing sample size with available spectroscopy will allow us to compute photo-z optimized for this type of source in the near future.

\begin{figure}
\centering
    \includegraphics[width=\linewidth]{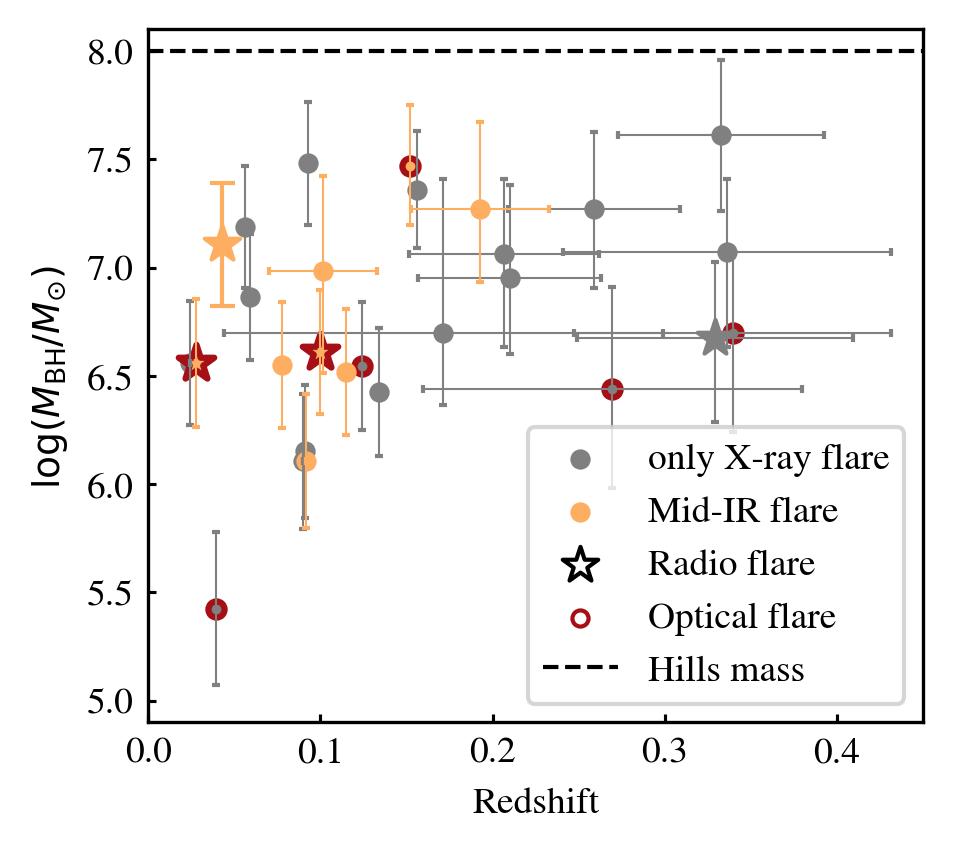}  
\caption{BH mass versus redshift  for the eROSITA TDE sample. Gray circles mark sources detected only in X-rays. Orange circles show sources that also had mid-IR flares, while red outlined circles show those that have optical flares. Sources with radio detections are denoted with stars. The dashed black line shows the Hills mass for a solar-type star and a non-spinning SMBH.}
\label{fig:bhmass}
\end{figure}

We used a scaling relation from \citet{reines2015} to estimate the BH masses from the host stellar masses, taking into account the 0.5dex scatter of the relation. The BH mass distribution versus redshift is shown in Fig.~\ref{fig:bhmass}, ranging from $\mathrm{log(M_{BH}/M_{\odot})}$=5.4 to 7.6 with a peak  $\mathrm{\approx6\times10^{6}M_{\odot}}$. This is below the Hills mass for the 1\,$\mathrm{M_{\odot}}$ star and a non-spinning BH \citep{1975Natur.254..295H}; it is thus consistent with the TDE interpretation. The lack of BHs with masses under $\mathrm{2\times10^{5}M_{\odot}}$ might be due to an increasingly smaller volume in which TDEs can still be detected around low-mass BHs \citep{Wevers_2017}, the BH occupation fraction in low-mass galaxies \citep{2016MNRAS.455..859S}, or inefficient circularization \citep{Dai_2015}.

It is evident from Fig.~\ref{fig:bhmass} that eROSITA TDEs accompanied by mid-IR flares are preferentially located at lower redshifts ($z<0.2$), with 60\,\% of mid-IR bright sources being below the X-ray luminosity median of the TDE sample ($\mathrm{L_{median}=2.4\times10^{43}ergs^{-1}}$). This might be associated with the obscuring envelope scenario (see Sect.~\ref{sec:multiwavelength_discussion}), which would make TDEs that are further away less likely to be discovered in X-rays due to obscuration. The lack of sources with mid-IR flares at higher redshifts also comes from the limited sensitivity of WISE. Although optically bright TDEs from the sample are more concentrated in the low BH mass regime, this is likely only related to selection effects (see Sect.~\ref{sec:opt_flares}). Indeed, comparing the BH mass distributions of our sample and optically selected samples \citep{yao2023,guolo2024systematic}, we found a very similar mean BH mass supporting the argument that the absence or detection of X-rays is independent of the BH mass.

\subsection{J1421: An off-nuclear TDE candidate}
\label{sec:j1421_imbh}

No SED modeling was performed for the source J1421. This TDE candidate differs from the rest of the sample as its optical counterpart was detected only as a transient in a single photometric band in two LS10 epochs (see Appendix~\ref{sec:j1421} for more details). The optical position is located $\approx 18$\,arcsec (21\,kpc) and $\approx 24$\,arcsec (28\,kpc) from the centers of two merging galaxies at z=0.06 (see Fig.~\ref{fig:J1421}). There is no separate underlying quiescent host detected at the position of the transient. An order-of-magnitude estimate of the corresponding BH mass can be obtained from the integrated energy released during the X-ray flare.

Integrating the X-ray flare  from eRASS1 to eRASS4 (see Fig.~\ref{fig:xray_lc}) provides a total energy release (in the X-ray band) of  $\mathrm{\approx 2.5\times 10^{50} erg}$. The emitted energy originates from accretion and can be roughly estimated as
\begin{equation}
   \mathrm{E_{tot} = \frac{f\eta GM_{BH}M_{*}}{R_{t}}},
\end{equation}
where $\mathrm{R_{t}}$ is the tidal radius, $\mathrm{\eta}$ is radiative efficiency, and f is the fraction of stellar mass accreted. Thus, solving for the BH mass for a Sun-like star with typical $\mathrm{\eta=0.1}$ and $\mathrm{f=0.5}$, we obtained $\mathrm{M_{BH}\approx5\times10^{4}M_{\odot}}$. Although the estimated mass may indicate that the X-ray flare could be linked to an off-nuclear TDE around an intermediate-mass black hole (IMBH), this calculation only provides a lower limit on mass. A better sampled X-ray light curve and multiwavelength observations would be required to test this hypothesis to estimate the total energy of the flare and provide better mass constraints.

Recent simulations showed that gravitational recoils or disruption of satellite galaxies caused by recent galaxy mergers could lead to a population of IMBHs out to $\geq 100$\,kpc \citep{2024MNRAS.tmp.2383U}. Such off-nuclear transients may represent promising IMBH TDE candidates in the future.

\section{Discussion}
\label{sec:discussion}

\subsection{X-ray Luminosity function and TDE occurrence rates}
\label{sec:xlf}

The 31 eROSITA candidates presented here form the largest systematically X-ray-selected sample of TDEs to date. The sample can be used to statistically describe the unobscured thermal population of TDEs with a canonical decaying light curve. We used the classical maximum observable volume ($\mathrm{1/V_{max}}$) approach \citep{1968ApJ...151..393S} to estimate the X-ray luminosity function (XLF). The calculation is based on the steps detailed in \citet{Grotova_2025}, which take into account eROSITA's inhomogeneous sensitivity, the selection criteria of the parent catalog eRO-ExTra, spectral shapes, luminosities, and redshifts of the sources. To calculate the XLF, we summed the derived $\mathrm{1/V_{max}}$ values of individual TDEs into seven equally spaced luminosity bins between $\mathrm{5\times10^{41}ergs^{-1}}$ and $\mathrm{1\times10^{45}ergs^{-1}}$. 

The uncertainty contribution for each spec-z source (i) was estimated as $\mathrm{(1/V_{max}^2)_i}$. For sources with large non-Gaussian photo-z errors (j), a more detailed approach was used to account for their $\mathrm{L_{X}}$ and $\mathrm{V_{max}}$ probability distributions. For each photo-z source, we performed Monte Carlo sampling for 100 values of $\mathrm{(V_{max})_j}$, which were randomly drafted from the photo-z probability distribution functions. Then, for each bin, we calculated the mean of $\mathrm{\Sigma(1/V_{max})_j}$ and its standard deviation $\mathrm{STD_{photo-z}}$ across all Monte Carlo realizations. The uncertainty, $\mathrm{\sqrt{STD_{photo-z}^2+\Sigma(1/V_{max})^2_i}}$, takes into account the spec-z and photo-z errors of sources in each bin.

As discussed in Sect.~\ref{sec:xray_props}, the selection is biased against sources peaking in eRASS1 due to the different typical rise and decay timescales and the primary selection cuts in the eRO-ExTra catalog. This results in us potentially missing a significant fraction of sources peaking in eRASS1, which did not decay in six months by a factor of more than four. Therefore, each source in the sample should be carefully weighted to account for this bias. For that, we identified 14 of 23 TDEs peaking in eRASS2, which would not have been selected if the amplitude and significance cuts were applied to the light-curve decay between eRASS2 and eRASS3. To account for this effect, we increased the weight of these 14 sources by a factor of two for the statistical completeness of the XLF. Thus, the final sample includes eight P1 sources with a weight of w=1, nine P2 sources with w=1, and 14 P2 sources with w=2. Accounting for the introduced weights, this results in 45 sources to be found per year. 

\begin{figure}
\centering
    \includegraphics[width=\linewidth]{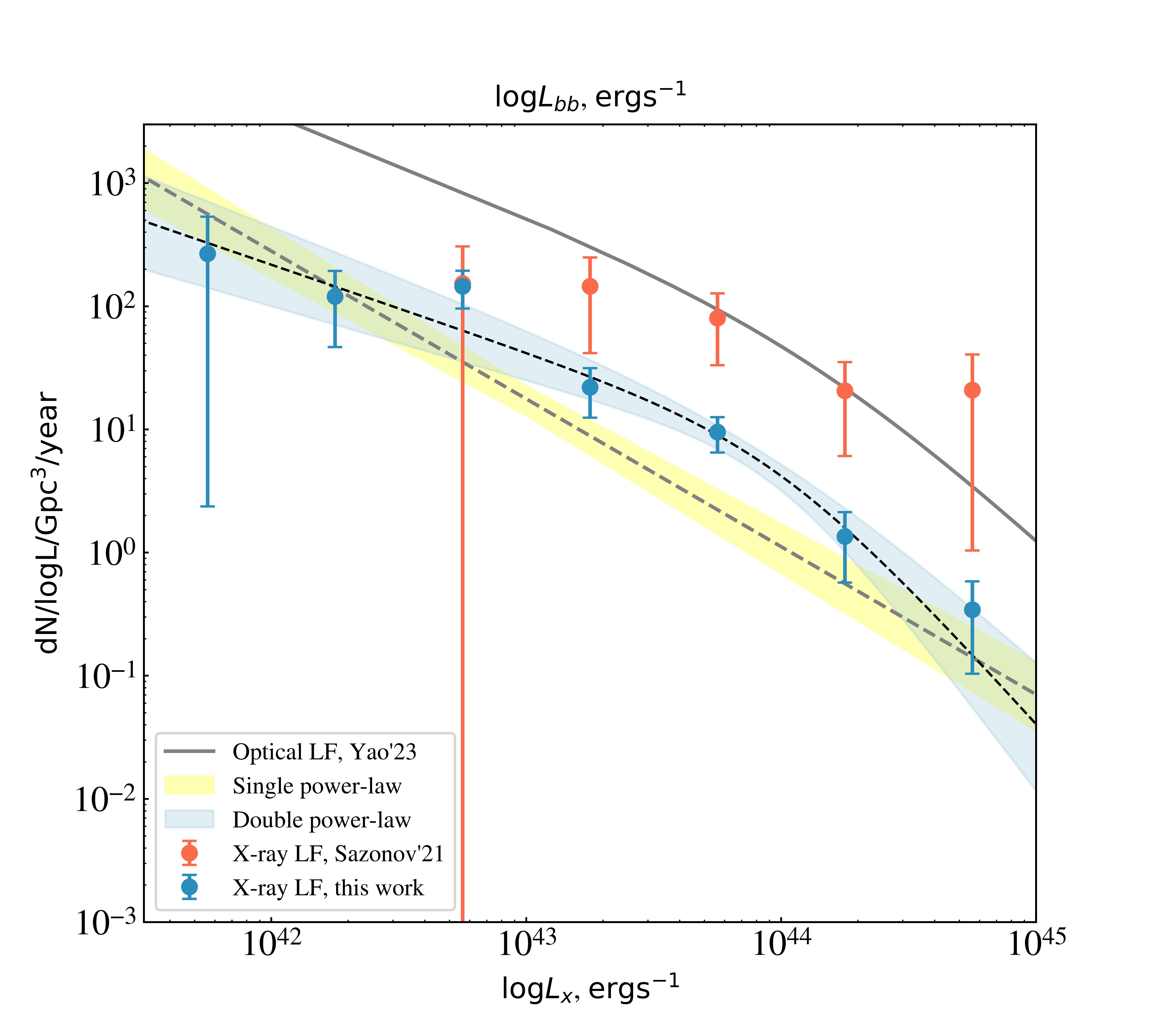}  
\caption{TDE XLF function in 0.2-6.0 keV. The blue points show the function obtained in this work, and the gray dashed line with the yellow region denotes its best fit with a power law (see Eq.\ref{eq:xlf_pow}); the blue region with a black dashed line shows the best fit with a double power law (see Eq.\ref{eq:double_xlf}). The orange points show the TDE XLF for eROSITA\_RU from \citet{Sazonov_2021}. The solid gray line shows the double power-law fit of a TDE optical luminosity function from \citet{yao2023}.}
\label{fig:xlf}
\end{figure}

Since the photo-z of J0847 is unreliable due to two missing bands, we did not use this source in the computation and instead implemented a correction factor of 45/43, as this source has a double weight. In addition, the function was multiplied by a factor of 2.6 to account for the whole sky area, since the sample was selected from the LS10 covered area (76\,\% = 1/1.3; see \citealt{Grotova_2025}) in the eROSITA\_DE hemisphere (1/2). Following the approach presented in \citet{Sazonov_2021}, we introduced an additional factor of three to account for sources missed due to the typical duration of a TDE. Overall, the total correction factor equals 8.16. The final XLF as well as other luminosity functions from \citet{Sazonov_2021} and \citet{yao2023} are presented in Fig.~\ref{fig:xlf}.

Fitting the XLF function with a single power law, we obtained: 

\begin{equation}
\label{eq:xlf_pow}
    \mathrm{\frac{dN_{TDE}}{dlogL_{X}dVdt} = B\times\left(\frac{L_{X}}{10^{43} ergs^{-1}}\right)^{-\beta} Mpc^{-3}yr^{-1}},
\end{equation}
where $\mathrm{B = (0.18\pm 0.05)\times 10^{-7}}$ and $\mathrm{\beta = 1.2\pm 0.08}$. The XLF function can be also fit with a double power-law function:

\begin{equation}
\label{eq:double_xlf}
\mathrm{
    f(L) = A \times \left( \left( \frac{L}{L_{\text{br}}} \right)^{-\alpha_{1}} + \left( \frac{L}{L_{\text{br}}} \right)^{-\alpha_{2}} \right)^{-1}Mpc^{-3}yr^{-1}},
\end{equation}
where $\mathrm{A = (0.8\pm 0.2)\times 10^{-7}}$, $\mathrm{\alpha_{1}= 0.7\pm0.1}$, and $\mathrm{\alpha_{2}= 2.3\pm0.4}$. Here, we fixed the break luminosity to $\mathrm{L_{br} = 1\times 10^{44}\, ergs^{-1}}$. This value corresponds to the Eddington-limiting prediction of an observable change of the XLF \citep{mummery21}, consistent with the significant decrease in detected sources with luminosities above this threshold. The value is also similar to the $\mathrm{L_{br}}$ found in previous studies (e.g., \citealt{yao2023,guolo2024systematic}). The double power-law significantly improved the fit to the data  ($\mathrm{\chi^2_{red}=0.82}$) compared to the single power-law fit ($\mathrm{\chi^2_{red}=3.05}$). 

By integrating the XLF (from Eq.~\ref{eq:double_xlf}) over the luminosity range of $\mathrm{5\times10^{41} < L_X < 1\times10^{45}\,erg s^{-1}}$, we estimated the average TDE volumetric rate to be $\mathrm{ (2.3^{+1.2}_{-0.9})\times10^{-7}\,Mpc^{-3} yr^{-1}}$. Using an estimation of the total galaxy volume density of $\mathrm{\approx 2\times 10^{-2}\,Mpc^{-3}}$ \citep{Bell_2003}, the TDE rate per year and per galaxy equals $\mathrm{\approx 1.2\times 10^{-5}\,yr^{-1} galaxy^{-1}}$.

\begin{table}[h]
    \centering
    \caption{Comparison of computed TDE rates.}

    \small
    \begin{tabular}{ccccc}
        \hline
        \hline
        Band & Rate & Reference & Sample \\
         & [$\mathrm{yr^{-1} gal^{-1}}$] & &size \\
        \hline
        X-ray & $9\times10^{-6}$ & \citet{Donley_2002} & 3 \\
        X-ray & $2\times10^{-4}$ & \citet{Esquej_2008} & 2  \\
        X-ray & $3\times10^{-5}$ & \citet{Khabibullin_sazonov} & 4 & \\
         X-ray & $1\times10^{-5}$ & \citet{Sazonov_2021} & 13 \\
         X-ray & $1.2\times10^{-5}$ & This work & 31 \\
        \hline
        Optical & $3.2\times10^{-5}$ & \citet{yao2023} & 33 \\
        Mid-IR & $2\times10^{-5}$ & \citet{masterson2024} &  12 \\
        \hline
    \end{tabular}
    
    \tablefoot{The top section shows the summary of X-ray TDE rates with step-by-step growing samples over two decades. The bottom section shows examples of recent TDE rates computed in other bands.}
    \label{tab:rate_summary}
\end{table}

Pre-eROSITA X-ray TDE rates were derived from ROSAT and XMM-Newton observations (see Table~\ref{tab:rate_summary}). The first XLF based on eROSITA was presented in \citet{Sazonov_2021} using 13 sources from eROSITA\_RU. This sample included eRASS2 sources out to z<0.6 with an amplitude larger than ten and without AGN signatures, and covered the brighter end of the luminosity function (L$_{\rm X}=\mathrm{10^{42.5}-10^{45} ergs^{-1}}$). Since we applied a lower amplitude cut in our approach, our XLF spans a lower range of luminosities down to $5\times 10^{41}$ erg $\mathrm{s^{-1}}$ and allowed us to probe the fainter end of the TDE population. Whereas both functions are consistent in the mid-luminosity regime ($\approx 10^{43}$ erg $\mathrm{s^{-1}}$), the slopes clearly diverge at higher luminosities. The discrepancy is due to different selection criteria used in defining the samples. The TDE sample used in this work contains purely canonical TDEs with decaying or flaring light curves, whereas TDEs in \citet{Sazonov_2021} include also sources ($\approx 30 \%$) brightening at later eRASSes. The exclusion of these sources from the latter sample makes the resulting function less steep and comparable with our XLF. Also selecting sources with amplitude larger than ten may contribute to the discovery of a flatter slope. The TDE rate of $\mathrm{(2.1\pm0.1)\times 10^{-7}Mpc^{-3}year^{-1}}$ presented in \citet{Sazonov_2021} is comparable with the one in this paper.

The slopes of our XLF broken double power-law fit are consistent with those obtained for optically selected TDE samples. In the X-ray sample of optically selected TDEs, \citet{guolo2024systematic} found $\mathrm{\alpha_{1,Guolo} = 0.96^{+0.21}_{-0.24}}$ and $\mathrm{\alpha_{2,Guolo} = 2.65^{+1.1}_{-0.9}}$, which agree with values presented in this study within the error bars. The slope values derived for the optical TDE luminosity function from \citet{yao2023} -- $\mathrm{\alpha_{1,Yao} = 0.84^{+0.30}_{-0.36}}$ and $\mathrm{\alpha_{2,Yao} = 1.93^{+0.32}_{-0.27}}$ -- also align with our results. In addition, the single power-law fit of TDE XLF reproduced using the first-principles relativistic disk theoretical approach in \citet{2024arXiv241017087M} corroborates our findings. There, the inferred XLF is scaled with luminosity as $\mathrm{\propto L_{X}^{-1.15}}$, which is very similar to the results obtained in our single power-law fit.

The normalization of our eROSITA TDE XLF is about a factor of ten lower (see Fig.~\ref{fig:xlf}) than that of optical TDEs from \citet{yao2023}. Here, it is important to note that the X-axes are different for both functions, and, therefore, they cannot be directly compared. We note that re-normalization of the two functions is challenging: it was demonstrated that for different TDE candidates, the difference between optical/UV and X-ray luminosities can lay within a very broad range of values with $\mathrm{L_{BB}/L_{X}\in(0.5,3000)}$ at early times and $\mathrm{L_{BB}/L_{X}\in(0.5,10)}$ at late times \citep{guolo2024systematic}. This results in the optical TDE rate being slightly higher than the X-ray rate. The similar rate difference was also found in previous X-ray studies (see Table~\ref{tab:rate_summary}). The possible reasons for lower X-ray rates compared to the theoretical estimates and the lack of a fainter TDE population might be 1) the fact that we did not include all possible X-ray TDE types in our canonical sample (see Sect.~\ref{sec:non-canonical}); 2) the lack of some X-ray TDEs due to obscuration or geometry effects (see Sect.~\ref{sec:multiwavelength_discussion}); 3) observational constraints of current X-ray all-sky surveys, which lead to the non-detection of fainter events; or 4) the fact that the primary selection criteria of the eRO-ExTra catalog did not pick up events with A<4. These factors should be explored in the future to confirm if optically and X-ray-selected populations of TDEs are indeed drawn from the same BH mass distribution \citep{2024arXiv241017087M}.

\subsection{Multiwavelength flares of TDE candidates}
\label{sec:flares_descr}

The presented TDE sample showed a wide variety of multiwavelength behaviors in the X-ray, optical, mid-IR, and radio bands. The diversity spans from TDE candidates detected only in X-rays to sources emitting in all four bands. The summary of the TDE multiwavelength flares is provided in Table~\ref{tab:tde_multicheck}, and the multiwavelength light curves are shown in  Appendix~\ref{appendix_multiwavelength}.

Mid-IR flares were detected both before and around the time of the X-ray peak. For example, in J0644 and J2108, mid-IR flares were found $\approx1500$ and $\approx1000$\,d before the X-ray detection. Also, in J0823, the mid-IR emission developed more than a year before the X-ray flare, consistent with the early optical peak. A detailed analysis of the multiwavelength properties of J0823 is provided in \citet{Malyali_AT2019avd}. J0439-65 and J2344-35 (more details in \citealt{homan2344}) peaked in the X-rays and mid-IR around the same time. However, the precise timing could not be constrained, given the sparsity of the eROSITA and NEOWISE data. Similarly, optical outbursts came noticeably earlier (e.g., J0318, J1102) or quasi-simultaneously (e.g., J0439) compared to X-ray peaks. Further discussion on the origin of mid-IR and optical flares in connection to X-ray flares is provided in Sect.~\ref{sec:multiwavelength_discussion}.

\begin{table}
\centering
\caption{Peak times for TDE candidates showing multiwavelength flares.}
\begin{tabular}{lrrrrr}
\hline
  \multicolumn{1}{c}{Name} &
    \multicolumn{1}{c}{LC} &

  \multicolumn{1}{c}{$\mathrm{\Delta t_{mid-IR}}$} &
  \multicolumn{1}{c}{$\mathrm{\Delta t_{optical}}$} &
  \multicolumn{1}{c}{$\mathrm{\Delta t_{radio}}$} \\
  \multicolumn{1}{c}{} &
    \multicolumn{1}{c}{type} &

  \multicolumn{1}{c}{[days]} &
  \multicolumn{1}{c}{[days]} &
  \multicolumn{1}{c}{[days]} 
 \\
\hline
\hline
  J0141& P1 &+175 & -& -\\
  J0318& P2& -&-631 &- \\
  J0439& P1& +191 & +145 &- \\
  J0608& P2& +146 &- & -\\
  J0615& P2& -12 &- &- \\
  J0644& P1& -1475 &- &-\\
  J0744& P2& -& -17&- \\
  J0823& P2&-2 & -615 &+79 \\
  J0847& P2& +177 & -& -\\
  J1102& P1& -& -381 & -\\
  J1305& P2& -& -&-55\\
  J1421& P2& -& -423 & -\\
  J1436& P1& +369 &- & -\\
  J1646& P2&- & -& -181\\
  J2108& P2& -920 & -& -539\\
  J2344& P2& +185 & -45 & +129\\
\hline\end{tabular}
\tablefoot{The $\Delta t$ in days is to the time of the brightest detection in the X-ray band. The P1 group includes sources with a single decline light curve peaking in eRASS1, and P2 includes those with a flare light curve peaking in eRASS2.}
\label{tab:tde_multicheck}
\end{table}

\subsection{Driving mechanisms of multiwavelength emission}
\label{sec:multiwavelength_discussion}

The order of the multiwavelength flares, their realtive time lag  or the total absence of observational features in certain bands have made it challenging to distinguish or unify X-ray, optical, and IR TDEs. Currently, two leading models are used explain the observed multiwavelength TDE scenarios.

Firstly, the obscuration model states that a newly formed TDE accretion disk is surrounded by an optically thick envelope formed from stellar debris. The envelope can reprocess soft X-rays from accretion onto the BH and produce optical and IR emission \citep{Roth_2016,Dai2018}. Particularly, the reprocessing layer can appear from winds from the accretion disk \citep{Strubbe_2009,Lodato_2010,Miller_2015} or stellar debris remaining in the vicinity of the black hole \citep{1997ApJ...489..573L, Guillochon_2014, metzger2016}. For X-ray fluxes high enough to sufficiently ionize the surrounding envelopes, the intrinsic X-ray emission can be observed simultaneously with the optical emission. This pattern was seen in sources J2344, and  J0439, where the X-ray and optical flares occured quasi-simultaneously within the observationally available cadence. On the other hand, a critically low intrinsic X-ray flux may lead X-rays to be completely absorbed. Indeed, numerous studies have identified optical TDEs without any X-ray signatures (e.g., \citealt{van_Velzen_2020}). Scenarios between these two extremes (due to lack of X-ray emission or envelope thickness) may lead to delayed X-ray flares, for example, as we saw in sources J1102 and J0318. Similarly, the mid-IR flares may precede X-ray flares due to dust reprocessing in the galaxy center.

The geometry of the surrounding envelope may also impact the observable properties of TDEs \citep{Guillochon_2014,Dai2018,parkinson22}. Depending on the viewing angle and the thickness of the obscuring material, we may either directly see the X-ray emission from the galactic center, or have a partially or fully obstructed view, looking perpendicularly through the envelope.  Several factors, for example, critical absorption or orientation of the surrounding envelope, can lead to extreme cases of X-ray non-detections. Whereas we still would detect TDE candidates in optical or IR bands, where the emission originates from reprocessing, this observational restriction can lead to underestimating TDE rates in X-rays.

Alternatively, TDEs can be described by the stream-stream collision model \citep{piran2015,Jiang2016,svirski2017}. In this scenario, the optical and X-ray photons originate in different locations. While X-rays appear in the case of efficient circularization of stellar debris and accretion onto the BH, optical emission can be produced earlier by shocks of stellar debris streams at larger distances. The effect can be even more prominent in the presence of magnetic stresses acting on the streams and strengthening self-crossing shocks \citep{bonnerot2017}. The stream-stream collision model describes TDEs with optical flares significantly preceding X-ray emission as well as the obscuring envelope model, by explaining delayed X-ray flares with inefficient circularization of the debris \citep{Gezari_2017,Liu_2022}. Some TDE candidates are better described with a stream-stream collision model. In particular, in the case of the multipeak optical light curve of J0823, the first peak was explained by a stream self-intersection, and the second peak was described by an accretion disk formation \citep{Malyali_AT2019avd,2022ApJ...928...63C}.

In the presented sample, only a fraction X-ray TDEs was detected in either mid-IR ($\approx 30\%$), optical ($\approx 20\%$), or both ($\approx 10\%$). The fractions should be considered only as lower limits due to significant observational constraints (see Sect.~\ref{sec:opt_flares}). For comparison, similar fractions were reported in the mid-IR population study by \citet{masterson2024} with $\approx 30\%$ sources detected in X-rays, and an optical population study by \citet{guolo2024systematic}, in which $\approx 40\%$ of optically selected TDEs were detected in X-rays. While these fractions are wrought with uncertainties from differences in multiwavelength observing strategies, they can help elucidate the geometry and physics of the underlying accretion flow.

\subsection{Non-canonical X-ray TDEs to be found}
\label{sec:non-canonical}

Whereas this work focused on the selection of canonical TDEs (with a soft spectrum, a single decline light curve, and a non-AGN quiescent host galaxy), future studies will explore more complex scnearios, for example, using the eRO-ExTra catalog or other surveys. Considering the non-canonical population will surely impact the rate and XLF prediction and help with the understanding of the unified picture of TDEs.

Firstly, recent discoveries (e.g. RX J133157.6-324319.7 \citealt{malyali_rep}; AT2018fyk \citealt{Wevers_2023}; eRASSt J045650.3-203750 \citealt{Liu_2023,Liu_2024}) have shown that X-ray TDEs can be repeating and pTDEs, where a star loses only a fraction of its mass during its initial encounter with the BH and survives, undergoing subsequent cycles of disruption during next orbital passes. The expected impact on the X-ray TDE rate is strengthened by the fact that the pTDE rate is predicted to be higher or comparable to full TDEs (e.g., \citealt{Bortolas}) since their encounter cross-sections are larger. The eRO-ExTra catalog contains potential pTDE candidates: 13 sources with a light-curve class other and one source with a brightening light curve have soft spectra at peak with $\mathrm{\Gamma>3}$. 

Secondly, the selection against sources with hard spectra may lead to missing some TDEs. For example, the accretion onto the SMBH during a TDE can lead to the formation of a corona \citep{guolo2024systematic, Liu_2023} while the TDE is still X-ray bright. Due to the time sampling of eROSITA observations (one visit every six months), such events may only be detected in the phase when the source is X-ray hard. Other TDE subclasses such as jetted TDEs \citep{2011Natur.476..421B,Saxton_2012,Bradley_Cenko_2012} were also found to have X-ray spectra at a peak with $\mathrm{\Gamma<2}$. In addition, intrinsically absorbed sources may also appear harder when being modeled with a simple power law. Similarly, a few of the 146 eRO-ExTra sources with a harder spectrum and a flaring or a decline in light curves might be missed TDE candidates. Due to factors such as sparse eROSITA coverage and unknown or low-luminosity AGN contamination, it is challenging to perform a systematic selection of a clean sample of non-canonical TDEs from eRO-ExTra. However, additional multiwavelength follow-up may help confirm the TDE origin of individual events.

Finally, TDEs are expected to occur not only in quiescent galaxies, but also in AGNs at higher rates \citep{Karas_2007}. However, it is challenging to identify a TDE in an AGN due to its low-contrast emission in comparison with the bright accretion disc and long-lived AGN variability caused by instabilities in the accretion flow \citep{komossa2015tidal}. Several X-ray and UV outbursts in AGNs were considered TDE candidates (e.g., \citealt{2020ApJ...898L...1R}). 

\section{Summary}
\label{sec:summary}
We present the sample of TDE candidates discovered in the eROSITA\_DE during its first and second all-sky surveys (Dec 2019 - Dec 2020). This is the largest systematically X-ray-selected sample of TDEs to date, which includes 31 TDE candidates. The sources were selected from the catalog of extragalactic non-AGN X-ray transients and variables eRO-ExTra \citep{Grotova_2025}, out to z<0.34, with a single decline or a flare X-ray light curve, a soft X-ray spectrum with $\Gamma>3,$ and no known long-term variability or AGN association. The sample includes 30 canonical TDEs and one off-nuclear TDE candidate. All sources remained X-ray soft throughout the evolution, except for J0439, of which the spectral hardening might be associated with the formation of the hot corona.

We computed TDE XLF, which is best fit with a double power law with a luminosity break at $\mathrm{10^{44}ergs^{-1}}$, in agreement with the Eddington-limiting prediction. Thus, we inferred the TDE rate of $\mathrm{\approx 1.2\times 10^{-5}\,yr^{-1} galaxy^{-1}}$. This is comparable with the prediction for the previous eROSITA X-ray TDE sample by \citet{Sazonov_2021} and slightly lower than the optical rate from \citet{yao2023}. The presented rate of X-ray-selected canonical TDEs does not consider missed events due to obscuration or geometry effects or nonstandard behavior (e.g. repeated or pTDEs). The SED modeling of host galaxies revealed an overdensity in the green valley, as was found in previous TDE studies. The BH masses of the canonical TDEs in the sample span from $\mathrm{log(M_{BH}/M_{\odot})}$=5.4 to 7.6.

A fraction of TDEs in the sample showed multiwavelength variability in the optical (20\%), mid-IR (30\%), and radio (15\%) bands, with optical and mid-IR outbursts either preceding or being quasi-simultaneous with the X-ray peak. We discuss the origin of multiwavelength flares in the context of the obscuring envelope model and the stream-stream collision model. Both scenarios can be used equally well to describe the observational data.

The exploration of TDEs with eROSITA should be continued to include events discovered in eRASS3--eRASS5. Furthermore, future studies should include non-canonical TDEs (e.g., from the eRO-ExTra catalog), which will improve the estimation of XLF and TDE rates and shed more light on our understanding of the TDE host population and multiwavelength properties. The methodology of the TDE identification presented in this work may also be applied in new X-ray surveys such as the Einstein Probe, which is expected to detect from several tens to hundreds of events per year \citep{2016IAUS..312...68Y, Yuan_2022}.

\begin{acknowledgements}
This work is based on data from eROSITA, the soft X-ray instrument aboard SRG, a joint Russian-German science mission supported by the Russian Space Agency (Roskosmos), in the interests of the Russian Academy of Sciences represented by its Space Research Institute (IKI), and the Deutsches Zentrum f\"ur Luft- und Raumfahrt (DLR). The SRG spacecraft was built by Lavochkin Association (NPOL) and its subcontractors and is operated by NPOL with support from the Max Planck Institute for Extraterrestrial Physics (MPE). The development and construction of the eROSITA X-ray instrument was led by MPE, with contributions from the Dr. Karl Remeis Observatory Bamberg  ECAP (FAU Erlangen-Nuernberg), the University of Hamburg Observatory, the Leibniz Institute for Astrophysics Potsdam (AIP), and the Institute for Astronomy and Astrophysics of the University of T\"ubingen, with the support of DLR and the Max Planck Society. The Argelander Institute for Astronomy of the University of Bonn and the Ludwig Maximilians Universit\"at Munich also participated in the science preparation for eROSITA. The eROSITA data shown here were processed using the eSASS/NRTA software system developed by the German eROSITA consortium.
      A. Malyali acknowledges support by DLR under the grant 50 QR 2110 (XMM\_NuTra). M. Krumpe acknowledges support by DLR under grant 50 OR 2307. The LCO observations have been made possible by the support of the Deutsche Forschungsgemeinschaft (DFG, German Research Foundation) under Germany’s Excellence Strategy-EXC-2094-390783311. This work was supported by the Australian government through the Australian Research Council’s Discovery Projects funding scheme (DP200102471). This paper includes data gathered with the 6.5 m Magellan Telescopes located at Las Campanas Observatory, Chile. A part of this work is based on observations made with the Southern African Large Telescope (SALT), with the Large Science Programmes on transients 2018-2-LSP-001 \& 2021-2-LSP-001 (PI: DAHB). Polish participation in SALT is funded by grant No. MEiN 2021/WK/01.This research has made use of the SIMBAD database,
operated at CDS, Strasbourg, France. This work has made use of data from the European Space Agency (ESA) mission
{\it Gaia} (\url{https://www.cosmos.esa.int/gaia}), processed by the {\it Gaia}
Data Processing and Analysis Consortium (DPAC,
\url{https://www.cosmos.esa.int/web/gaia/dpac/consortium}). Funding for the DPAC
has been provided by national institutions, in particular the institutions
participating in the {\it Gaia} Multilateral Agreement. The ZTF forced-photometry service was funded under the Heising-Simons Foundation grant 12540303 (PI: Graham).      
\end{acknowledgements}

\bibliographystyle{aa}
\bibliography{bibliography}

\begin{appendix}
\onecolumn

\section{X-ray properties of the TDE sample}
\begin{figure}
\centering
    \includegraphics[width=0.95\linewidth]{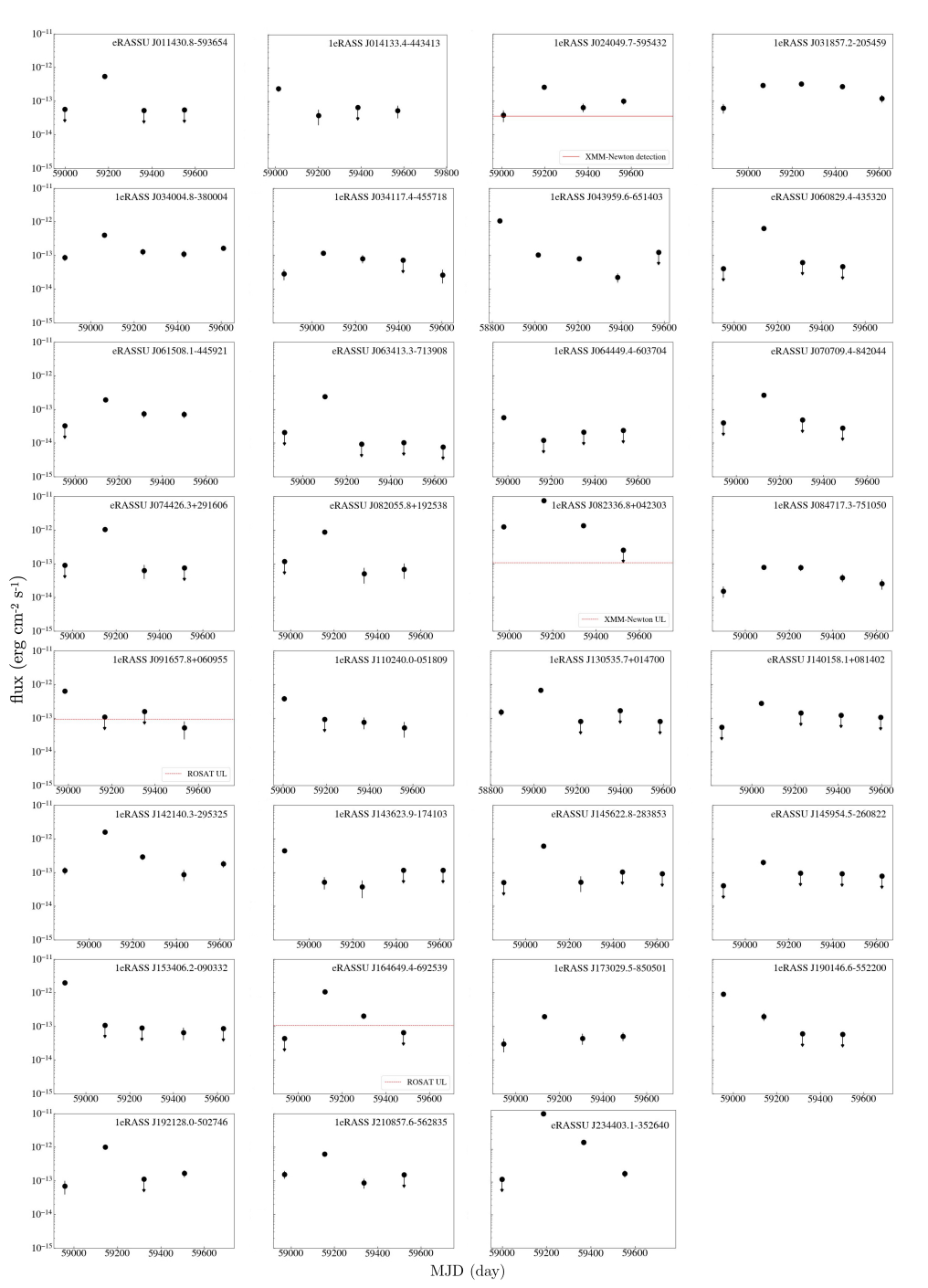}  
\caption{eRASS1-4(5) eROSITA X-ray light curves for the TDE sample in 0.2-2.3 keV. The red dashed and solid lines denote archival ULs and detections, respectively.}
\label{fig:xray_lc}
\end{figure}

The eROSITA spectral modeling results for eRASS1-4(5) are provided in Table~\ref{tab:modeling_xray}. The eROSITA eRASS1-4(5) lightcurves are shown in Fig.~\ref{fig:xray_lc}.

\begin{table}[h!]
\centering
\caption{Spectral modeling results of TDE candidates for eRASS1-eRASS5 in keV.}
\begin{tabular}{lrrrrr}
\hline
\hline
  \multicolumn{1}{c}{ERO\_NAME} &
  \multicolumn{1}{c}{$\mathrm{kT_{eRASS1}}$} &
  \multicolumn{1}{c}{$\mathrm{kT_{eRASS2}}$} &
  \multicolumn{1}{c}{$\mathrm{kT_{eRASS3}}$} &
  \multicolumn{1}{c}{$\mathrm{kT_{eRASS4}}$} &
  \multicolumn{1}{c}{$\mathrm{kT_{eRASS5}}$}  \\
\hline
  eRASSU J011430.8-593654 & -   & 0.1$\pm$ 0.01   &  -  & -   &  - \\
  
  1eRASS J014133.4-443413 & 0.08 $\pm$ 0.01 & - &  -  & - &  - \\
  
  1eRASS J024049.7-595432 & -& 0.20 $\pm$ 0.02 & 0.28 $\pm$ 0.08 & 0.27 $\pm$ 0.06 &  - \\
  
  1eRASS J031857.2-205459  & 0.05 $\pm$ 0.01 & 0.09 $\pm$ 0.01 & 0.08 $\pm$ 0.01 & 0.08 $\pm$ 0.01 & 0.07 $\pm$ 0.01\\
  1eRASS J034004.8-380004  & 0.24 $\pm$ 0.07 & 0.17 $\pm$ 0.01 & 0.29 $\pm$ 0.05 & 0.58 $\pm$ 0.22 & 0.2 $\pm$ 0.03\\
  1eRASS J034117.4-455718  & - & 0.17 $\pm$ 0.04 & 0.12 $\pm$ 0.04 &  -  & -\\
  1eRASS J043959.6-651403  & 0.08 $\pm$ 0.002 & 0.21 $\pm$ 0.02 & 0.14 $\pm$ 0.03 & 0.35 $\pm$ 0.27 &  - \\
  eRASSU J060829.4-435320  & -   & 0.11 $\pm$ 0.01 &   -   &   - & -\\
  eRASSU J061508.1-445921  & -   & 0.14 $\pm$ 0.02 & 0.12 $\pm$ 0.03 & 0.14 $\pm$ 0.03 &  - \\
  eRASSU J063413.3-713908  & -   & 0.12 $\pm$ 0.0 &  -  & - &- \\
  1eRASS J064449.4-603704  & 0.08 $\pm$ 0.01 &   -   &  -  & -  & -\\
  eRASSU J070709.4-842044  & -   & 0.14 $\pm$ 0.01 &  -  &  -    & -\\
  eRASSU J074426.3+291606 & -   & 0.1 $\pm$ 0.01 & - &     - &- \\
  
  eRASSU J082055.8+192538  & -   & 0.08 $\pm$ 0.01 & - & - &-   \\
  1eRASS J082336.8+042303  & 0.09 $\pm$ 0.01 & 0.09 $\pm$ 0.0 & 0.08 $\pm$ 0.01 &  -  &- \\
  1eRASS J084717.3-751050  & - & 0.32 $\pm$ 0.06 & 0.5 $\pm$ 0.18 & 0.33 $\pm$ 0.11 & -\\
  1eRASS J091657.8+060955  & 0.06 $\pm$ 0.01 &    -  & - &-  & -  \\
  1eRASS J110240.0-051809  & 0.08 $\pm$ 0.01 &  -  & - & - &  - \\
  1eRASS J130535.7+014700  & 0.09 $\pm$ 0.02 & 0.11 $\pm$ 0.01 &  -  &  -  & -\\
  eRASSU J140158.1+081402  & -   & 0.06 $\pm$ 0.01 &  -  & -   &- \\
  1eRASS J142140.3-295325  & 0.17 $\pm$ 0.03 & 0.17 $\pm$ 0.01 & 0.16 $\pm$ 0.02 & - & 0.16 $\pm$ 0.03\\
  1eRASS J143623.9-174103  & 0.1 $\pm$ 0.01 & - & - &  - &- \\
  eRASSU J145622.8-283853  &  -  & 0.09 $\pm$ 0.01 & - & - & - \\

  eRASSU J145954.5-260822  &  -  & 0.18 $\pm$ 0.04 &  -  & -  &- \\
  1eRASS J153406.2-090332  & 0.12 $\pm$ 0.0 &  -  &   - & - &  - \\
  eRASSU J164649.4-692539  & -   & 0.1 $\pm$ 0.0 & 0.13 $\pm$ 0.02 &  -&  - \\
  1eRASS J173029.5-850501  & - & 0.24 $\pm$ 0.03 & - & 0.52 $\pm$ 0.16 & -  \\
  1eRASS J190146.6-552200  & 0.09 $\pm$ 0.01 & 0.06 $\pm$ 0.01 &  -  & - &  - \\
  1eRASS J192128.0-502746  & - & 0.2 $\pm$ 0.02 &  -  & 0.29 $\pm$ 0.07 & -  \\
  1eRASS J210857.6-562835  & 0.19 $\pm$ 0.05 & 0.13 $\pm$ 0.01 & 0.16 $\pm$ 0.08 &  -  & -\\
  eRASSU J234403.1-352640  &   -& 0.08 $\pm$ 0.002 & 0.05 $\pm$ 0.003 & 0.04 $\pm$ 0.01 & -  \\
\hline\end{tabular}
\tablefoot{The modeling was performed for all detections with sufficient counts with detection likelihood DET\_LIKE>15.}
\label{tab:modeling_xray}
\end{table}
\FloatBarrier

\newpage

\section{J1421 - an optical transient in LS10}
\label{sec:j1421}

\begin{figure}[h!]
\centering
    \includegraphics[width=0.5\linewidth]{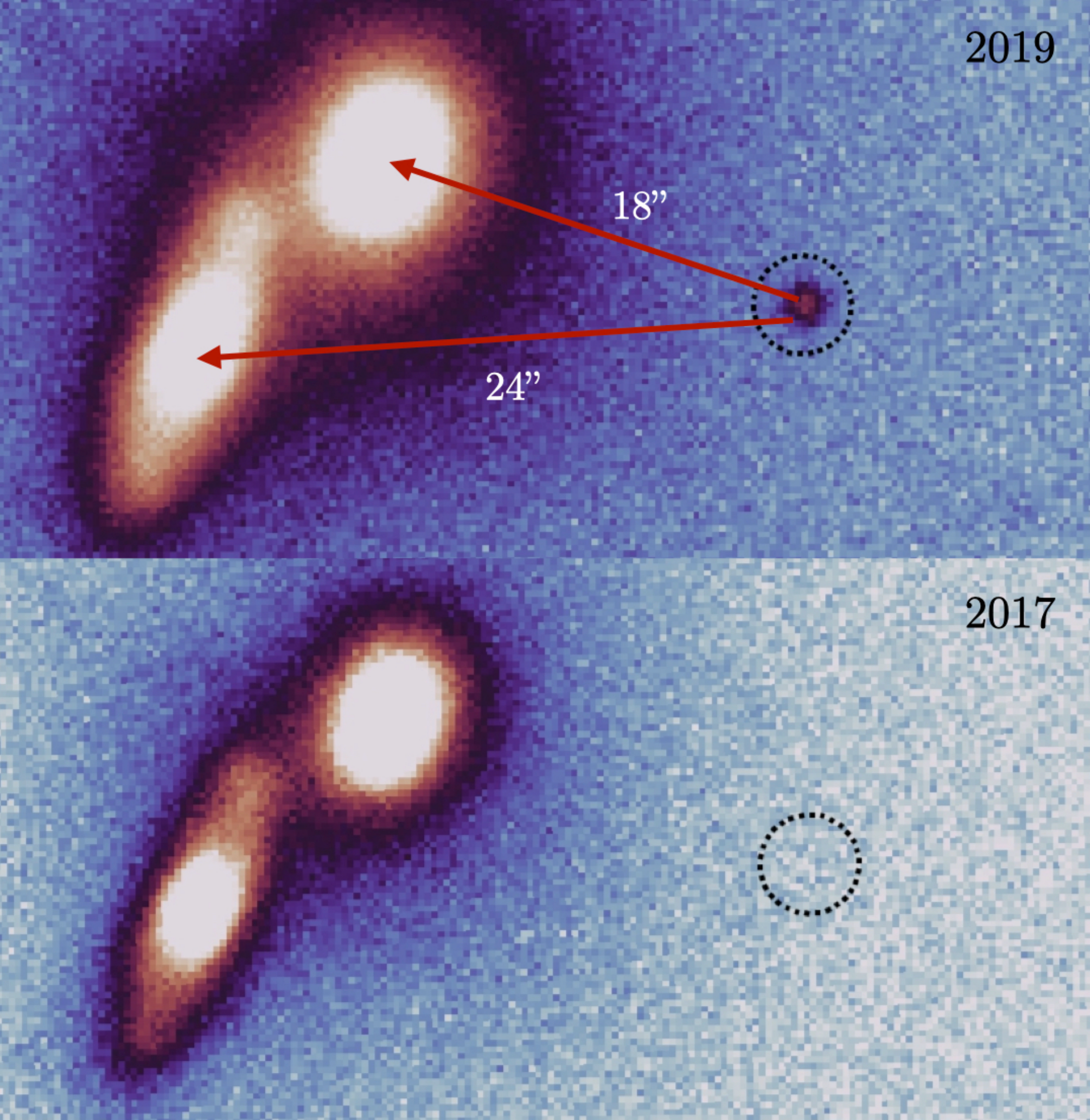}  
\caption{LS10 images of the optical transient at the location of J1421. Top: the detection of the transient in {\it g} band in 2019 (MJD 58643). The distances from the centers of two merging galaxies at z=0.06 are $\approx 18$\,arcsec (21\,kpc) and $\approx 24$\,arcsec (28\,kpc), respectively. Bottom: the non-detection of the source in {\it g} band in 2017 (MJD 57850).}
\label{fig:J1421}
\end{figure}

In the LS10 scanning strategy, each sky area is not observed simultaneously in {\it griz} optical bands but rather is covered at different times. In cases of drastic variability, such as the appearance of a new source, images taken at different bands can be used to identify transients. The inspection of the LS10 images revealed that the optical counterpart was detected only in the {\it g} band and only in two consecutive epochs in 2019 (MJD 58643.14 and 58643.16, see Fig.~\ref{fig:J1421}). It was not detected in any other bands or epochs between 2015 and 2021\footnote{\url{https://www.legacysurvey.org/viewer/exposures/?ra=215.4154&dec=-29.8895&layer=ls-dr10}}. The optical emission of 19.21 mag appeared at least 14 months before the eROSITA peak (MJD 59064) in eRASS2 and has faded to a non-detection during the X-ray decline phase, which was confirmed by a total absence of the source during a follow-up observation with NTT in March 2021.

The source is likely extragalactic and physically associated with a pair of nearby merging galaxies. At their common redshift of z = 0.06, the distance between the center of the closest galaxy and the g-band source is only $\approx 20$ kpc. The luminosity ($\mathrm{L_{X}\approx 1.4\times 10^{43} erg s^{-1}}$), soft X-ray spectrum ($\mathrm{kT \approx 170 eV}$), and flaring light curve would be consistent with an off-nuclear TDE. As the optical flare was detected approximately one year before the X-ray peak, this would require the X-ray emission to either be from a repeating event or a delayed late-time response. Other non-TDE extragalactic origins cannot be excluded. However, we point out that soft X-ray spectra would be atypical for standard AGNs and/or fast blue optical transients (e.g., \citealt{Bright_2022}). The long duration of the X-ray event (J1421 was detected in all five eRASSes) and much earlier optical detection make a Galactic stellar flare origin very unlikely.

This example shows how our view of optical and X-ray TDE candidates depends on the scanning strategy, sensitivity of the instruments and precise timing of observations. If the LS10 observation coincidentally was not taken during the optical brightening phase, the source would be classified as X-ray bright only. Moreover, there would be no evidence to assume that the source is not nuclear.

\newpage

\newpage
\section{Multiwavelength flares of TDEs}
\label{appendix_multiwavelength}

\begin{figure}
\centering
    \includegraphics[width=0.91\linewidth]{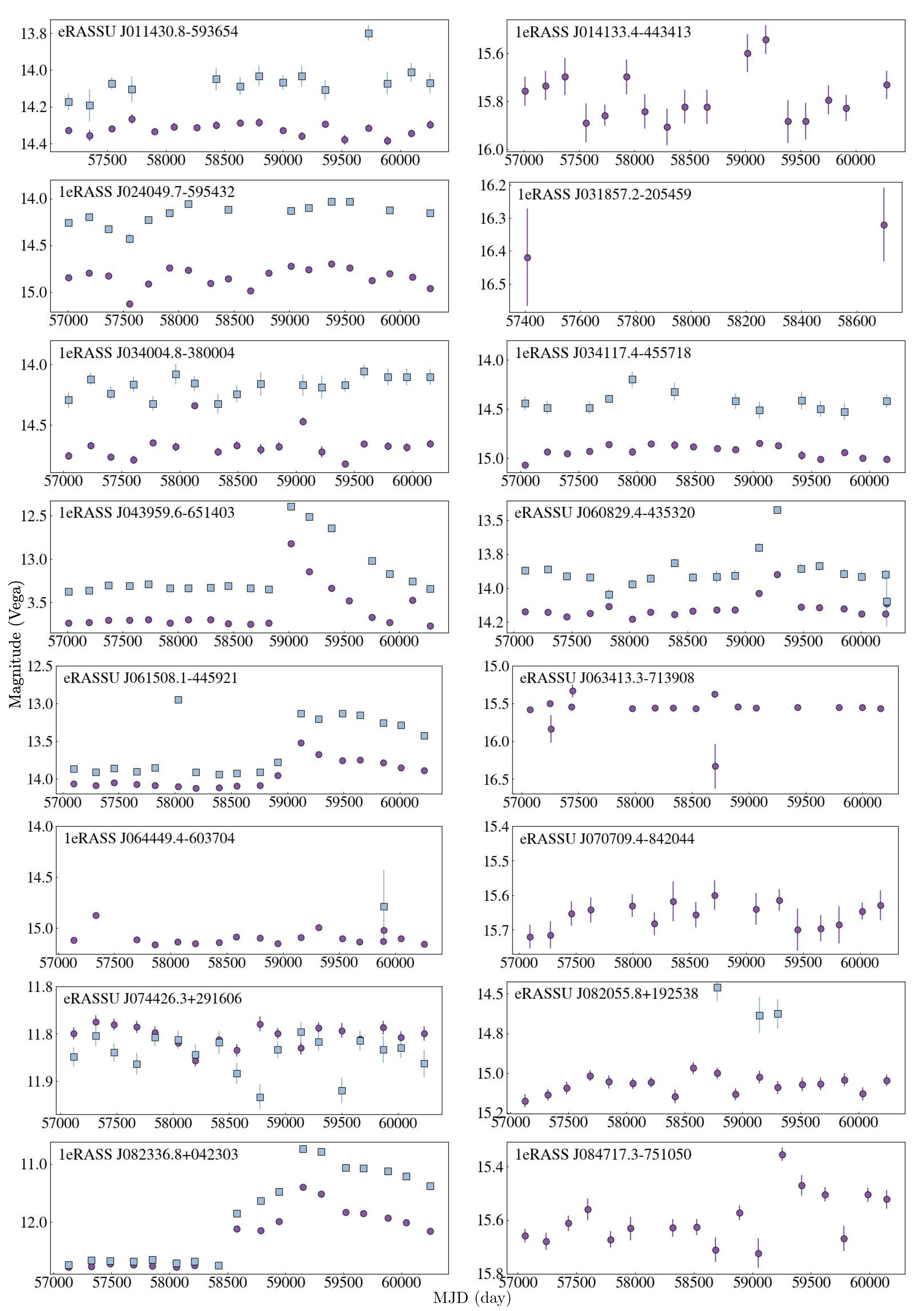}  
\caption{NEOWISE W1 and W2 light curves, binned in ten-day intervals, for the TDE sample. Purple circles denote W1 data points, and blue squares show W2 data points.}
\label{fig:wise_lc1}
\end{figure}

\begin{figure}
\centering
    \includegraphics[width=0.91\linewidth]{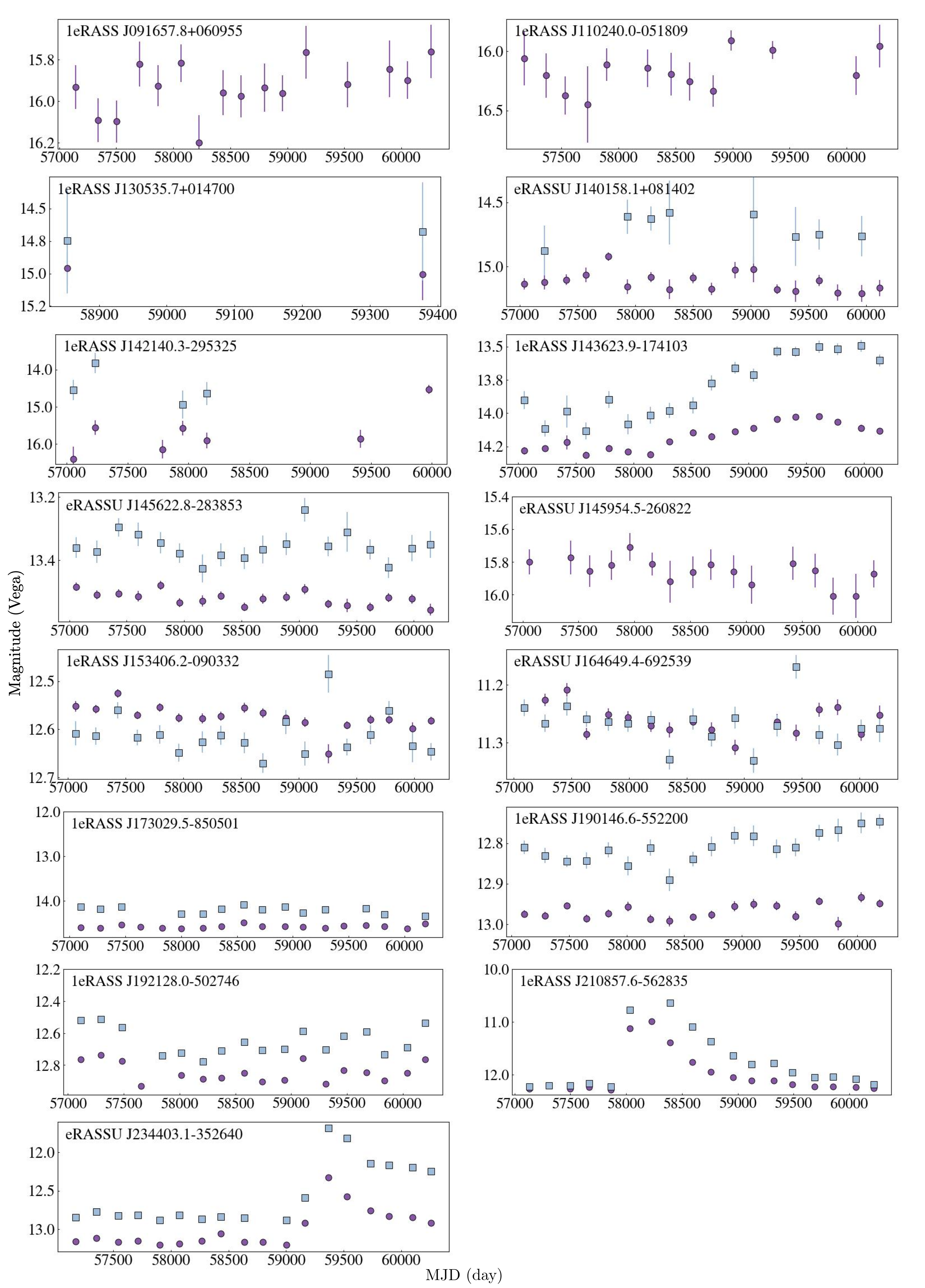}  
\caption{The same figure as Fig.~\ref{fig:wise_lc1}.}
\label{fig:wise_lc2}
\end{figure}

\begin{figure}
    \centering
    \begin{minipage}[t]{0.33\textwidth}
        \centering
        \includegraphics[width=\textwidth]{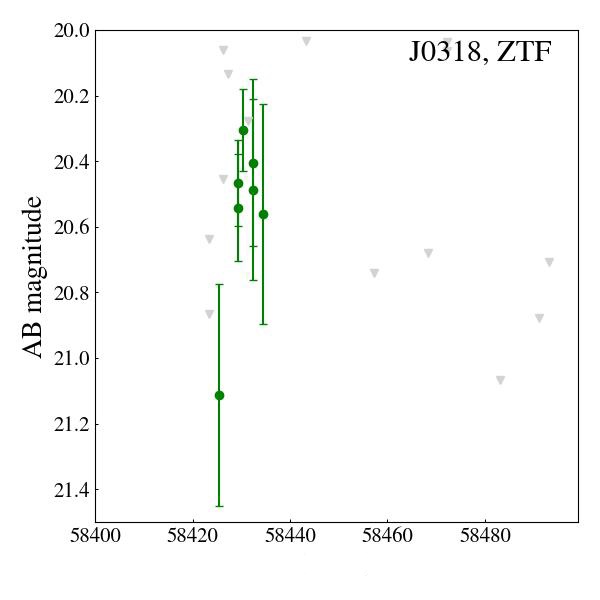}
    \end{minipage}
    \begin{minipage}[t]{0.33\textwidth}
        \centering
        \includegraphics[width=\textwidth]{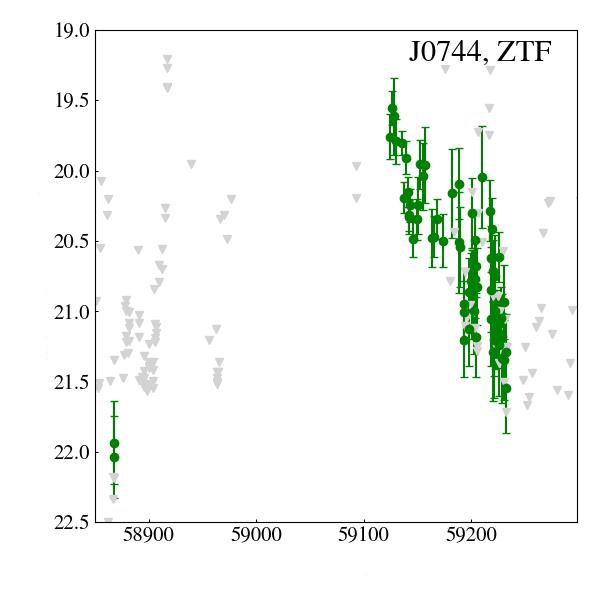}
    \end{minipage}
    \begin{minipage}[t]{0.33\textwidth}
        \centering
        \includegraphics[width=\textwidth]{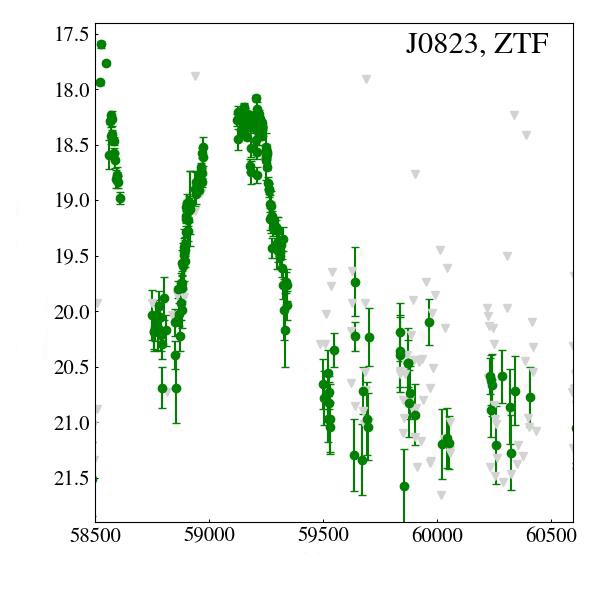}
    \end{minipage}

    \vspace{-0.2cm}


    \begin{minipage}[t]{0.33\textwidth}
        \centering
        \includegraphics[width=\textwidth]{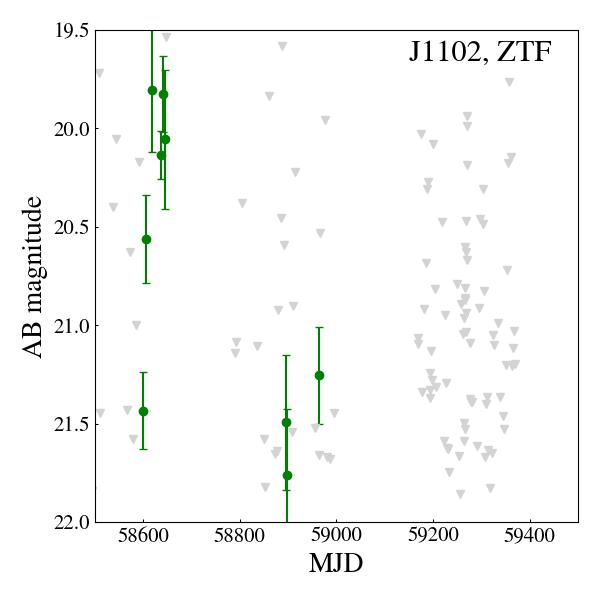}
    \end{minipage}
    \begin{minipage}[t]{0.33\textwidth}
        \centering
        \includegraphics[width=\textwidth]{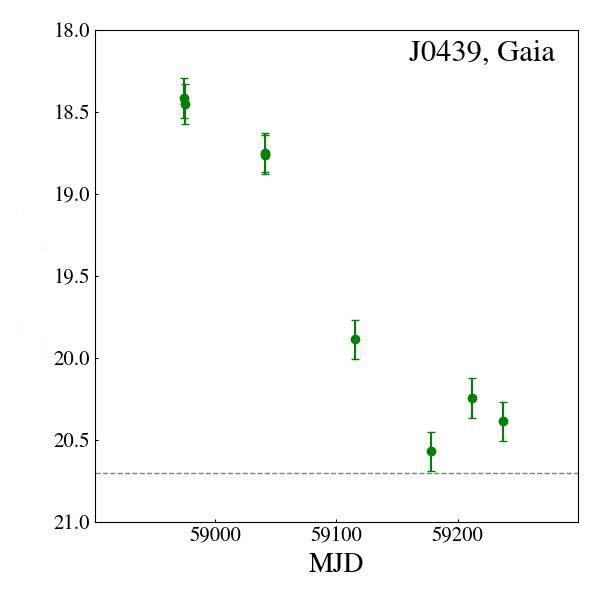}
    \end{minipage}
    \begin{minipage}[t]{0.33\textwidth}
        \centering
        \includegraphics[width=\textwidth]{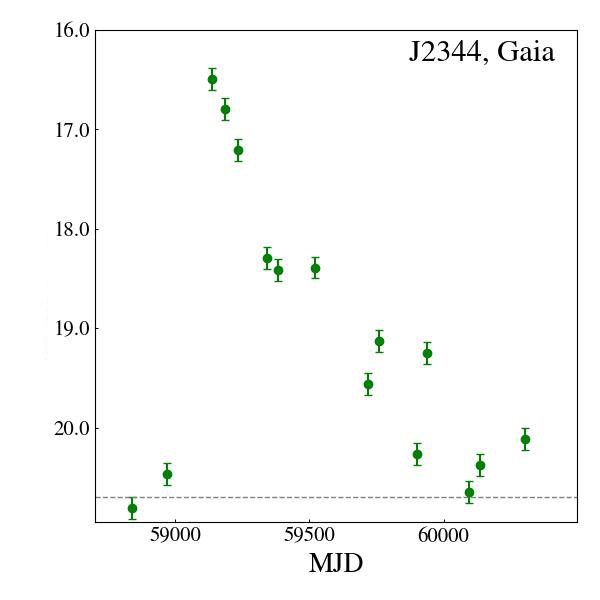}
    \end{minipage}

    \caption{Host-subtracted light curves of optical flares of six sources from the TDE sample in {\it g} band. Gray triangles denote upper limits in ZTF light curves. Gray dashed lines show the Gaia detection limit of 20.7 mag \citep{2021A&A...652A..76H}.}
    \label{fig:optical_lc_ztf_gaia}
\end{figure}

This section presents additional mid-IR and optical properties and light curves for the TDE sample. The mid-IR non-host-subtracted NEOWISE light curves for all 31 TDEs are shown in Fig.~\ref{fig:wise_lc1} and ~\ref{fig:wise_lc2}. To filter the reliable W1, W2 data points we applied quality indicator factor $\mathrm{qi\_fact>0.9}$. The optical host-subtracted light curves for sources showing a significant flare (see Table~\ref{tab:optical_flares}) are shown in Fig.~\ref{fig:optical_lc_ztf_gaia}. The summary of used optical data and calculated ATLAS ULs is presented in Table~\ref{tab:optical_lc_ul}.

\begin{table}[h!]
\centering
\caption{Optical magnitudes at peak for the TDE sample.}
\begin{tabular}{lcll}
\hline
\hline
  \multicolumn{1}{c}{Name} &
  \multicolumn{1}{c}{Survey} &
  \multicolumn{1}{c}{Peak magnitude} &
  \multicolumn{1}{c}{Transient ID} \\

  \multicolumn{1}{c}{ } &
  \multicolumn{1}{c}{} &
  \multicolumn{1}{c}{[AB mag]} &
  \multicolumn{1}{c}{ } \\
\hline

  J0114-59 & ATLAS & $19.0^{\triangledown}$ & \\
  J0141-44 & ATLAS & $18.7^{\triangledown}$ & \\
  J0240-59 & ATLAS & $19.0^{\triangledown}$ & \\
  J0318-20 & ZTF & $20.3\pm 0.1$ & \\
  J0340-38 & ATLAS & $19.1^{\triangledown}$ & \\
  J0341-45 & ATLAS & $18.9^{\triangledown}$ & \\
  J0439-65 & Gaia &$18.41 \pm 0.12$ & Gaia20cdq/AT2020jgh \\
  J0608-43 & ATLAS & $18.8^{\triangledown}$ & \\
  J0615-44 & ATLAS & $19.0^{\triangledown}$& \\
  J0634-71 & ATLAS & $18.8^{\triangledown}$& \\
  J0644-60 & ATLAS & $19.0^{\triangledown}$ & \\
  J0707-84 & ATLAS & $18.8^{\triangledown}$ & \\
  J0744+29 & ZTF & $19.6 \pm 0.1$ & ZTF20acgrymn \\
  J0820+19 & ATLAS & $19.0^{\triangledown}$ & \\
  J0823+04 & ZTF & $17.13 \pm 0.04$ & AT2019avd \\
  J0847-75 & ATLAS & $18.9^{\triangledown}$ & \\
  J0916+06 & ATLAS & $18.9^{\triangledown}$ & \\
  J1102-05 & ZTF & $19.8 \pm 0.3$ & \\
  J1305+01 & ATLAS & $19.0^{\triangledown}$ & \\

  J1401+08 & ATLAS & $19.0^{\triangledown}$ & \\
  J1421-29 & ATLAS & $18.3^{\triangledown}$ & \\
  J1436-17 & ATLAS & $19.0^{\triangledown}$ & \\
  J1456-28 & ATLAS & $18.2^{\triangledown}$ & \\
  J1459-26 & ATLAS & $18.7^{\triangledown}$ & \\
  J1534-09 & ATLAS & $18.9^{\triangledown}$ & \\
  J1646-69 & ATLAS & $18.6^{\triangledown}$ & \\
  J1730-85 & ATLAS & $18.6^{\triangledown}$ & \\
  J1901-55 & ATLAS & $18.8^{\triangledown}$ & \\
  J1921-50 & ATLAS & $18.9^{\triangledown}$ & \\
  J2108-56 & ATLAS & $18.8^{\triangledown}$ & \\
  J2344-35 & Gaia & $16.50 \pm 0.11$ & Gaia20eub/AT2020wjw \\
\hline
\end{tabular}

\tablefoot{Detections are provided in the {\it g} band with $1\sigma$ errors at the peak time of the optical flare. Triangles represent $\mathrm{5\sigma}$ ATLAS upper limits in the {\it o}-band for sources without detected optical flares.}
\label{tab:optical_lc_ul}
\end{table}

\newpage

\onecolumn
\section{SED modeling}

This section presents SED modeling results in Table~\ref{tab:sed_results}.

\begin{table}[h!]
\centering
\caption{SED modeling results for 28 sources from eROSITA TDE sample.}
\begin{tabular}{lrrr}

\hline
\hline
  \multicolumn{1}{c}{Name} &
  \multicolumn{1}{c}{Mass} &
  \multicolumn{1}{c}{mfrac} &
  \multicolumn{1}{c}{$^{0,0}u-r$} \\

\multicolumn{1}{c}{} &
  \multicolumn{1}{c}{$\mathrm{\times10^{10}M_{\odot}}$} &
  \multicolumn{1}{c}{} &
  \multicolumn{1}{c}{[mag]} \\
\hline  

  J0240-59 & $6.88^{+3.8}_{-3.76}$ & 0.62 & $1.68^{+0.33}_{-0.25}$ \\
  J0340-38 & $5.09^{+5.59}_{-1.97}$ & 0.66 & $1.63^{+0.22}_{-0.27}$ \\
  J0341-45 & $6.52^{+3.64}_{-3.47}$ & 0.67 & $1.31^{+0.32}_{-0.18}$ \\
  J0615-44 & $5.84^{+6.56}_{-3.32}$ & 0.62 & $2.13^{+0.10}_{-0.14}$ \\
  J0707-84 & $3.18^{+9.79}_{-1.11}$ & 0.61 & $2.08^{+0.10}_{-0.79}$ \\
  J1102-05 & $3.01^{+2.05}_{-1.63}$ & 0.64 & $1.58^{+0.42}_{-0.33}$ \\
  J1305+01 & $3.0^{+1.76}_{-1.36}$ & 0.61 & $1.9^{+0.29}_{-0.37}$ \\
  J1730-85 & $22.5^{+14.93}_{-8.48}$ & 0.63 & $1.95^{+0.31}_{-0.25}$ \\
  J0114-59 & $13.43^{+1.91}_{-1.74}$ & 0.61 & $2.31^{+0.05}_{-0.05}$ \\
  J0141-44 & $0.88^{+0.17}_{-0.16}$ & 0.6 & $2.1^{+0.06}_{-0.06}$ \\
  J0318-20 & $1.77^{+2.14}_{-0.97}$ & 0.61 & $1.63^{+0.31}_{-0.48}$ \\
  J0439-65 & $16.35^{+4.0}_{-2.53}$ & 0.64 & $2.21^{+0.06}_{-0.06}$ \\
  J0608-43 & $2.17^{+0.49}_{-0.32}$ & 0.64 & $1.78^{+0.11}_{-0.11}$ \\
  J0634-71 & $1.72^{+0.39}_{-0.3}$ & 0.62 & $1.98^{+0.06}_{-0.06}$ \\
  J0644-60 & $2.12^{+0.44}_{-0.33}$ & 0.61 & $2.0^{+0.06}_{-0.06}$ \\
  J0744+29 & $0.19^{+0.05}_{-0.03}$ & 0.62 & $2.07^{+0.08}_{-0.08}$ \\
  J0820+19 & $2.26^{+0.51}_{-0.41}$ & 0.61 & $2.07^{+0.08}_{-0.08}$ \\
  J0823+04 & $2.29^{+0.57}_{-0.42}$ & 0.62 & $2.25^{+0.08}_{-0.08}$ \\
  J0916+06 & $1.02^{+0.19}_{-0.17}$ & 0.57 & $2.11^{+0.05}_{-0.05}$ \\
  J1401+08 & $10.65^{+7.37}_{-4.37}$ & 0.63 & $1.82^{+0.26}_{-0.28}$ \\
  J1436-17 & $10.92^{+10.67}_{-4.0}$ & 0.61 & $2.14^{+0.16}_{-0.18}$ \\
  J1456-28 & $17.55^{+4.86}_{-3.81}$ & 0.61 & $2.12^{+0.06}_{-0.06}$ \\
  J1459-26 & $0.88^{+0.18}_{-0.16}$ & 0.6 & $2.11^{+0.06}_{-0.06}$ \\
  J1534-09 & $2.31^{+0.32}_{-0.29}$ & 0.61 & $2.15^{+0.07}_{-0.07}$ \\
  J1901-55 & $4.51^{+1.2}_{-0.94}$ & 0.61 & $2.07^{+0.08}_{-0.08}$ \\
  J1921-50 & $9.1^{+2.26}_{-1.72}$ & 0.62 & $2.15^{+0.07}_{-0.07}$ \\
  J2108-56 & $7.85^{+1.84}_{-1.56}$ & 0.6 & $1.97^{+0.10}_{-0.10}$ \\
  J2344-35 & $2.41^{+0.47}_{-0.33}$ & 0.65 & $2.07^{+0.08}_{-0.08}$ \\
\hline
\end{tabular}

\tablefoot{The mass column lists host stellar masses. The surviving mass, used in this work, was obtained by multiplying the total mass by the coefficient mfrac. The $^{0,0}u-r$ represents the rest frame colors.}
\label{tab:sed_results}
\end{table}

\end{appendix}

\end{document}